\documentclass[12pt,preprint]{aastex}

\shorttitle{[C II] line emission from z$\sim$6 quasars}
\shortauthors{Wang et al.}

\begin{document}


\title{Star Formation and Gas Kinematics of Quasar Host Galaxies at z$\sim$6: New insights from ALMA}

\author{Ran Wang\altaffilmark{1,2,14},
Jeff Wagg\altaffilmark{3},
Chris L. Carilli\altaffilmark{1},
Fabian Walter\altaffilmark{4},
Lindley Lentati\altaffilmark{5},
Xiaohui Fan\altaffilmark{2},
Dominik A. Riechers\altaffilmark{6},
Frank Bertoldi\altaffilmark{7},
Desika Narayanan\altaffilmark{2},
Michael A. Strauss\altaffilmark{8},
Pierre Cox\altaffilmark{9},
Alain Omont\altaffilmark{10},
Karl M. Menten\altaffilmark{11},
Kirsten K. Knudsen\altaffilmark{12},
Roberto Neri\altaffilmark{9},
Linhua Jiang\altaffilmark{13}}
\altaffiltext{1}{National Radio Astronomy Observatory, PO Box 0, Socorro, NM, USA 87801}
\altaffiltext{2}{Steward Observatory, University of Arizona, 933 N Cherry Ave., Tucson, AZ, 85721, USA}
\altaffiltext{3}{European Southern Observatory, Alonso de C\'ordova 3107, Vitacura, Casilla 19001, Santiago 19, Chile}
\altaffiltext{4}{Max-Planck-Institute for Astronomy, K$\rm \ddot o$nigsstuhl 17, 69117 Heidelberg, Germany}
\altaffiltext{5}{Astrophysics Group, Cavendish Laboratory, JJ Thomson Avenue, Cambridge, CB3 0HE, UK}
\altaffiltext{6}{Astronomy Department, Cornell University, 220 Space Sciences Building, Ithaca, NY 14853, USA}
\altaffiltext{7}{Argelander-Institut f$\rm \ddot u$r Astronomie, University of Bonn, Auf dem H$\rm \ddot u$gel 71, 53121 Bonn, Germany}
\altaffiltext{8}{Department of Astrophysical Sciences, Princeton University, Princeton, NJ, USA, 08544}
\altaffiltext{9}{Institute de Radioastronomie Millimetrique, St. Martin d'Heres, F-38406, France}
\altaffiltext{10}{Institut d'Astrophysique de Paris, CNRS and Universite Pierre et Marie Curie, Paris, France}
\altaffiltext{11}{Max-Planck-Institut f$\rm \ddot u$r Radioastronomie, Auf dem H$\rm \ddot u$gel 71, 53121 Bonn, Germany}
\altaffiltext{12}{Chalmers University of Technology, Department of Earth and Space Sciences, Onsala Space Observatory, SE-43992 Onsala, Sweden}
\altaffiltext{13}{School of Earth and Space Exploration, Arizona State University, Tempe, AZ 85287-1504, USA, Hubble Fellow}
\altaffiltext{14}{Jansky Fellow}


\begin{abstract}
We present Atacama Large Millimeter/submillimeter Array (ALMA) observations 
of the [C II] 158 $\mu$m fine structure line and dust continuum emission 
from the host galaxies of five redshift 6 quasars. 
We also report complementary observations of 250 GHz dust continuum and CO (6-5) line emission from
the z=6.00 quasar SDSS J231038.88+185519.7 using the IRAM facilities. 
The ALMA observations were carried 
out in the extended array at $\rm 0.7''$ resolution. We have detected 
the line and dust continuum in all five objects. The derived [C II] line luminosities 
are 1.6$\rm \times10^{9}$ to 8.7$\rm \times10^{9}\,L_{\odot}$ and the [C II]-to-FIR 
luminosity ratios are $\rm 2.9-5.1\times10^{-4}$, which is comparable to the 
values found in other high-redshift quasar-starburst systems and local ultra-luminous infrared 
galaxies. The sources are marginally resolved and the intrinsic source sizes (major axis FWHM) 
are constrained to be $\rm 0.3''$ to $\rm 0.6''$ (i.e., 1.7 to 3.5 kpc) for 
the [C II] line emission and $\rm 0.2''$ to $\rm 0.4''$ (i.e., 1.2 to 2.3 kpc) for the continuum.
These measurements indicate that there is vigorous star formation over the central few kpc 
in the quasar host galaxies. The ALMA observations also constrain the 
dynamical properties of the star-forming gas in the nuclear region. 
The intensity-weighted velocity maps of 
three sources show clear velocity gradients. 
Such velocity gradients are consistent with a rotating, gravitationally bound gas component, 
although they are not uniquely interpreted as such.  
Under the simplifying assumption of rotation, the implied dynamical masses 
within the [C II]-emitting regions are of order $\rm 10^{10}$ 
to $\rm 10^{11}\,M_{\odot}$. Given these estimates, the mass ratios between the SMBHs and 
the spheroidal bulge are an order of magnitude higher than the mean value found in 
local spheroidal galaxies, which is in agreement with results from previous 
CO observations of high redshift quasars. 
\end{abstract}





\section{Introduction}

Quasars at redshift 6 and higher provide a unique sample 
to study the formation of the first supermassive black holes (SMBHs) and their 
host galaxies at the epoch of cosmic reionization. 
There are currently more than sixty quasars known at z$\sim$6, 
selected from large optical and near-infrared 
surveys (e.g. the Sloan Digital Sky Survey [SDSS], \citealp{fan06,jiang08,jiang09}; the Canada-France 
High Redshift Quasar Survey, \citealp{willott07,willott09,willott10}, the UKIDSS Large Area Survey [ULAS], 
\citealp{mortlock09,mortlock11}, and the Panoramic Survey Telescope and 
Rapid Response System [Pan-STARRS], \citealp{morganson12}) 
with optical z-band magnitudes from 18.8 to 24.4 
and inferred SMBH masses from $\rm 10^{8}$ to 
a few $\rm 10^{9}\,M_{\odot}$ \citep{kurk07,jiang07,willott10,derosa11}. 
The formation of $\rm 10^{9}\,M_{\odot}$ SMBHs at the highest redshift 
suggests fast black hole accretion and significant SMBH-galaxy evolution within 1 Gyr after the Big Bang \citep{li07,narayanan08,di12}. 
The co-evolution of the first SMBHs and their host galaxies   
are studied with observations of dust continuum, molecular CO, 
and [C II] line emission at submillimeter and millimeter [(sub)mm] 
wavelengths (e.g., \citealp{priddey01,walter03,maiolino05}).   
Dust heated by the UV photon from young, massive stars present strong 
thermal FIR continuum emission. Due to the negative K-correction \citep{blain93}, 
observations of dust continuum emission at (sub)mm wavelengths provide an efficient 
way to search for star forming activity at high redshifts \citep{omont96,omont03,
priddey03,bertoldi03a,wang07}.
Molecular gas traced by CO transitions provides the requisite fuel for 
star formation \citep{narayanan08,carilli13b}, and the [C II] 158$\mu$m fine 
structure line emission 
is a principle interstellar coolant, which probes the star 
forming-powered photodissociation regions (PDR) and interstellar medium in 
these earliest quasar host galaxies \citep{walter09,delooze11,gallerani12,wagg12,venemans12,carilli13b}.

Dust and high-order molecular CO line emission in the host galaxies of 
the z$\sim$6 quasars have been searched for using the Max 
Planck Millimeter Bolometer Array (MAMBO) on the IRAM-30m 
telescope \citep{bertoldi03a,petric03,wang07,wang08,wang11a,omont13}, 
the Submillimeter Common User Bolometer Array (SCUBA) on the 
James Clerk Maxwell Telescope\citep{priddey03,robson04,priddey08}, 
and the IRAM Plateau de Bure interferometer (PdBI, 
\citealp{bertoldi03b,walter03,carilli07,wang10,wang11a}). 
The MAMBO survey of SDSS z$\sim$6 quasars at 250 GHz (typical 1$\sigma$ errors of 0.6 mJy) found that
about 30\% of them show strong continuum emission from dust at a temperature 
of 40 to 60 K, with FIR 
luminosities of a few $\rm 10^{12}$ to $\rm 10^{13}\,L_{\odot}$ and inferred dust masses 
of a few $\rm 10^{8}\,M_{\odot}$ \citep{priddey01,bertoldi03a,petric03,wang11a}.  
Most of the FIR-luminous z$\sim$6 quasars also show bright 
CO line emission, implying that there is $\rm \sim10^{10}\,M_{\odot}$ of molecular
gas in the quasar host galaxies \citep{walter03,carilli07,wang10,wang11b}.
The strong FIR continuum and molecular CO line emission suggest that there is
active star formation in the host galaxies of these FIR luminous z$\sim$6 quasars.
The star formation rates inferred from the FIR luminosities range from about a 
few hundred to one thousand $\rm M_{\odot}\,yr^{-1}$ \citep{bertoldi03a,wang08,wang11a}. 

The CO detections from the FIR-luminous z$\sim$6 quasars 
also provide constraints on the spatial distribution of molecular 
gas and the dynamical masses of the quasar host galaxies \citep{walter04,wang11b}. 
Deep CO line imaging of one of the most                
FIR-luminous quasars, SDSS J114816.64+525150.3 at z=6.42 (hereafter J1148+5251), 
using the Very Large Array (VLA) 
and the PdBI \citep{walter04,riechers09} at sub-arcsecond 
resolution reveals a CO source size of about 3.6 kpc$\rm \times$1.4 kpc (FWHM)
in the central region of the quasar host galaxies.
Recent JVLA observations of the CO (2-1) line emission from 
another FIR-luminous quasar, SDSS J092721.82+200123.7 at z=5.77, 
at 2$''$ resolution has constrained the source size 
to be about 10 kpc \citep{wang11b}. The line widths of the CO-detected 
z$\sim$6 quasars range widely from 160 to 860 $\rm km\,s^{-1}$.
The host galaxy dynamical masses (within the CO emitting region) 
estimated from the CO line widths and the sizes of the bright detections indicate a median 
SMBH-bulge mass ratio about one order of magnitude higher than the present-day
value \citep{walter04,wang10}. 

Strong [C II] 158 $\mu$m line emission was also 
detected from the host galaxies of the FIR and CO-luminous quasars 
at z$\sim$6 \citep{maiolino05,maiolino09,willott13}. 
The PdBI image of [C II] line of J1148+5251 shows two components. The narrow-line  
component with a line width of $\rm 345\,km\,s^{-1}$ and 
a spatial extension of $\rm FWHM\sim1.5''$ (8 kpc) traces the distributed star 
formation in the quasar host galaxy \citep{maiolino12} and the 
compact core of this component within the central $\rm \sim0.3''$ 
region suggest a very high peak star formation rate surface density 
of $\rm \sim1000\,M_{\odot}\,yr^{-1}\,kpc^{-2}$ \citep{walter09}.
The broad-line component shows a 
line width of FWHM$\sim$2000 $\rm km\,s^{-1}$, distributed over $\sim$16 kpc, suggesting gas outflow 
driven by the central luminous quasar \citep{maiolino12,valiante12}.
The detection of dust continuum, CO, and [C II] line emission in the quasar host 
galaxies at z$\sim$6 suggest an early phase of SMBH-galaxy evolution, in which 
the SMBHs are accreting at their Eddington limit and the quasar stellar bulges  
are still accumulating their mass via massive star formation.  
The interferometer observations of J1148+5251 have demonstrated that the [C II] line emission 
is an ideal tracer of star-forming activity and gas dynamics in the nuclear region 
of the starburst quasar host galaxies at the highest redshift. 

The Atacama Large Millimeter/submillimeter Array (ALMA) in the 
early science phase provides a wide 
frequency coverage for observations of fine structure line emission at high redshift. 
It also provides the required sensitivity and allows detections 
of line emission with peak flux density of a few mJy in only one hour of observing time.
Additionally, the spatial resolution of $\sim$0.7$''$ in Band 6 and 7 
in the extended configuration (with a maximum baseline of $\sim$400 m) 
can measure or tightly constrain the distribution of the gas component from high redshift quasar host galaxies. 
Thus, the Cycle 0 phase of ALMA opens a unique opportunity to expand the observation of [C II] line emission to a large sample of 
quasar-starburst systems at the highest redshift.
In this paper, we present ALMA observations of the [C II] line emission from five 
FIR luminous quasars at z$\sim$6. We also report IRAM observations of the 250 GHz 
dust continuum and CO (6-5) line emission from the quasar SDSS J231038.88+185519.7 
at z=6.00 (Fan et al. 2013 in prep.). 
We describe the observations in Section 2, present the results 
in Section 3, discuss the star forming activity and gas dynamics based on 
the [C II] line detections in Section 4, and summarize the results in 
Section 5. A $\rm \Lambda$ CDM cosmology with $\rm H_{0}=71km\ s^{-1}\ Mpc^{-1}$, $\rm
\Omega_{M}=0.27$ and $\rm \Omega_{\Lambda}=0.73$ is adopted throughout this
paper \citep{spergel07}.

\section{Observations}

\subsection{IRAM observations of SDSS J231038.88+185519.7}

The quasar SDSS J231038.88+185519.7 (hereafter J2310+1855) 
is a broad absorption line quasar discovered in the SDSS 
(Fan et al. 2013, in prep.). The redshift measured 
with the quasar rest-frame UV line emission is z=6.00$\pm$0.03. 
It is one of the optically brightest sources among 
the known z$\sim$6 quasars with rest frame 1450 $\rm \AA$ magnitude of $\rm m_{1450}=19.3$. 
We observed the 250 GHz dust continuum from 
this object with the MAMBO-II 117-element array on the IRAM 30-m telescope 
\citep{kreysa98} in the winter of 2010-2011. 
We adopted the standard on-off photometry mode with a chopping 
rate of 2 Hz by $\rm 32''$ in azimuth. The total observing time was 100 minutes 
with 50 minutes on source. The data 
were reduced with the MOPSIC pipeline \citep{zylka98}. We reached a 
1$\rm\sigma$ sensitivity of 0.63 mJy and the source is detected at 8.29 mJy($\rm \sim13\sigma$), making it 
the brightest 250 GHz detection among all the known z$\sim$6 quasars. 

We then searched for the CO (6-5) line emission from this object using the PdBI. The observation 
was carried out in 2011 in D configuration with a
synthesized beam size (FWHM) of $\rm 5.4''\times3.9''$. We spent 3 hours on source using 
the new wide band correlator WideX, with a bandwidth of 3.6 GHz in dual polarization. The phase was 
checked with the quasar 3C 454.3 and the flux calibration uncertainty is 5\%. 
The data were reduced with the IRAM GILDAS software package 
(Guilloteau \& Lucas 2000). We detected the line at 11$\sigma$ with a typical 
rms of $\rm 0.5\,mJy\,Beam^{-1}$ per 100 $\rm km\,s^{-1}$ channel (Table 1 and Figure 1). We also detected 
the 99GHz dust continuum at 8$\sigma$ with a continuum 1$\sigma$ sensitivity of $\rm 0.05\,mJy\,Beam^{-1}$.

\subsection{ALMA observations}

To study the [C II] 158 $\mu$m line emission and the underlying continuum with ALMA, 
We selected J2310+1855 and another four z$\sim$6 quasars that were 
detected in bright millimeter dust continuum and CO (6-5) 
line emission (\citealp{petric03,priddey03,priddey08,wang08,wang10,wang11a}; Fan et al. 2013 in 
prep.) and have Declination of $\rm \delta\leq$ +20$^{\circ}$. 
We list the sources and measurements from previous MAMBO observations in Table 1.
The ALMA observations were carried out in Cycle 0 
in 2012 with 14 to 21 12-m diameter antennas in the extended array with baselines from 36 m to 400 m. 
The typical synthesized beam size (FWHM) is $\rm 0.7''$, corresponding 
to 4 kpc at z$\sim$6. We observed the [C II] 158 $\mu$m (1900.5369 GHz) 
line emission in Band 6 or 7, with one 2 GHz window centered at the line 
frequency and the other three 2 GHz windows observing the dust continuum. 
The correlator channel width is 15.625 MHz, or 16$\sim$18 $\rm km\,s^{-1}$ 
at the observed line frequency. The phase stability was checked every $\sim$8 minutes by 
observing nearby phase calibrators (e.g. bright quasars). We spent 50 to 90 minutes on source for 
each target and reached 1$\sigma$ spectral sensitivities of 0.4 to 
0.7 $\rm mJy\,Beam^{-1}$ per 62.5 MHz re-binned channel after 
continuum subtraction. The continuum 
sensitivities are 0.04 to 0.09 $\rm mJy\,Beam^{-1}$. 

We reduced the data using the CASA package\footnote{http://casa.nrao.edu/}. 
The flux scale was calibrated in four of the sources 
by observing Titan, Neptune, and Uranus, and the calibration uncertainties are 
better than 15\%. For the source J1044$-$0125, no primary amplitude calibrator was 
observed. The flux scale was determined 
with 3C 273 based on the ALMA and SMA observations in March  
and April 2012 at 225 GHz and 343 GHz. Considering the 
uncertainties in the flux densities and spectral index of the calibrator, 
we estimate a total flux calibration error of 20\% for the observation of J1044$-$0125. 
We derive the line frequencies in the kinematic local standard of rest (LSRK) 
frame to compare the results to previous PdBI CO (6-5) line observations. 

\section{Results}

We detected strong [C II] line and dust continuum emission 
in the host galaxies of all the five z$\sim$6 quasars. 
The line emission of the five objects and the continuum emission from four  
of them (the exception is J1044$-$0125) are marginally resolved. 
The line intensity-weighted velocity maps of
J0129$-$0035, J1319+0950, and J2310+1855 show clear velocity
gradients (Figure 2). 
We fit the sizes of the line or continuum sources 
using the IMFIT task in CASA, which performs synthesized beam deconvolution and 
2-dimensional Gaussian fitting to the images.  
The resulting deconvolved full width at half maximum (FWHM) major axis sizes 
are 0.3$''$ to $\rm 0.6''$ (1.7 to 3.5 kpc) for the line and 
0.2$''$ to $\rm 0.4''$ (1.2 to 2.3 kpc) for the 
dust continuum. We list the fitted parameters for each object below. 
However, we emphasize that, due to the limited spatial resolution, 
these measurements should be considered as tight constraints on 
the spatial extent of the sources, rather than accurate source morphology measurements. 
Interferometer imaging at higher resolution is required to finally 
determine the source morphology. 
To measure the line center, FWHM line width, and flux, we 
subtract the underlying dust continuum in the U$-$V 
plane, calculate the line spectra by integrating the intensity over the 
line-emitting region in each channel, and fit the spectra with a 
Gaussian profile. 
The [C II] results are summarized in Table 2, and the line and continuum maps are presented 
in Figure 2, and 3.  

{\bf SDSS J2310+1855}. This object is the brightest detection among all the z$\sim$6 
quasars from our MAMBO observations. The 250 GHz continuum flux 
density is $\rm 8.29\pm0.63\,mJy$. Our PdBI observation of the CO(6-5) line emission 
shows a line flux of $\rm 1.52\pm0.13\,Jy\,km\,s^{-1}$, which is also by far the 
strongest CO line flux. The redshift 
and FWHM measured with the CO (6-5) line are $\rm z_{CO}=6.0025\pm0.0007$ 
and $\rm FWHM_{CO}=456\pm64\,km\,s^{-1}$. The dust continuum at 99 GHz has also been 
detected in the line-free channels with a flux density of 0.40$\pm$0.05 mJy (Figure 1).

We detect the [C II] line emission with a flux 
which is again more than twice stronger than the four other sources. 
The derived FWHM source size, after deconvolving  
the  $\rm 0.72''\times0.51''$ synthesized beam, 
is $\rm (0.56''\pm0.03'')\times(0.39''\pm0.04'')$ with a position angle (PA) of $\rm 142^{\circ}\pm10^{\circ}$ 
east of north. 
The line intensity-weighted velocity 
map derived with the line-emitting channels 
indicates a velocity gradient  
from Southwest to Northeast (See the right panels of Figure 2). 
The 263 GHz continuum flux density imaged with 
the line-free windows and fitted to a 2-D Gaussian is $\rm 8.91\pm0.08\,mJy$, 
which is in very good agreement with the MAMBO value. 
This constrains the (deconvolved) continuum source size to 
be $\rm (0.25''\pm0.02'')\times(0.20''\pm0.02'')$ with $\rm PA=162^{\circ}\pm18^{\circ}$. 

{\bf J1319+0950}. This quasar was discovered in the UKIRT Infrared
Deep Sky Survey (UKIDSS) and has $\rm m_{1450}=19.65$ \citep{mortlock09}. 
The [C II] line emission is imaged by ALMA with a synthesized beam size of $\rm 0.69''\times0.49''$.  
The 2-D Gaussian fit suggests line emission with a 
deconvolved source size of $\rm (0.57''\pm0.07'')\times(0.32''\pm0.15'')$ oriented 
at $\rm PA=28^{\circ}\pm18^{\circ}$. 
The line velocity map shows a clear gradient from the Southwest to the Northeast. 
We plot the [C II] line channel map of this object in Figure 3. The map shows a clear position 
shift of the [C II] line peak (by $\rm \sim0.4''$) from the $\rm +222\,km\,s^{-1}$ channel to 
the $\rm -200\,km\,s^{-1}$ channel.   
The 258 GHz continuum flux density is $\rm 5.23\pm0.10\,mJy$ and the deconvolved continuum 
source size is $\rm (0.39''\pm0.02'')\times(0.34''\pm0.03'')$ with $\rm PA=121^{\circ}\pm148^{\circ}$.

{\bf J2054$-$0005}. The quasar was selected from SDSS stripe 82 
with $\rm m_{1450}=20.60$, i.e. about one magnitude fainter than the 
objects discovered from the SDSS main survey \citep{jiang08}. We observed 
the [C II] line emission at $\rm 0.64''\times0.58''$ resolution. 
The deconvolved FWHM source size is 
$\rm (0.35''\pm0.04'')\times(0.32''\pm0.05'')$ 
and $\rm PA=91^{\circ}\pm173^{\circ}$.
The 262 GHz continuum flux density measured in the line-free windows 
is $\rm 2.98\pm0.05\,mJy$, and the deconvolved continuum source size is 
$\rm (0.27''\pm0.03'')\times(0.26''\pm0.03'')$ with $\rm PA=168^{\circ}\pm68^{\circ}$.

{\bf J0129$-$0035}. This quasar was discovered in SDSS Stripe 82 
with $\rm m_{1450}=22.16$ \citep{jiang09}. It is the 
faintest optical source among the MAMBO-detected z$\sim$6 quasars, but previous millimeter 
observations found bright 250 GHz 
dust continuum and CO (6-5) line emission from this object \citep{wang11a}. 
The [C II] line was observed at $\rm 0.57''\times0.49''$ resolution. 
The deconvolved source size is estimated to be $\rm (0.41''\pm0.06'')\times(0.23''\pm0.12'')$ 
with $\rm PA=12^{\circ}\pm22^{\circ}$. 
A velocity gradient is found from the Southeast to the Northwest. 
A 2-D Gaussian fit to the 
continuum image gives a 287 GHz flux density of $\rm 2.57\pm0.06\,mJy$, with a 
deconvolved continuum source size of $\rm (0.23''\pm0.04'')\times(0.14''\pm0.07'')$ 
at $\rm PA=115^{\circ}\pm154^{\circ}$.

{\bf J1044$-$0125}. This is another broad absorption 
line quasar at z$\sim$6 discovered from the SDSS main 
survey with $\rm m_{1450}=19.2$ \citep{fan00}. 
A deep near-IR spectrum of the quasar C IV line emission 
of this object yields a SMBH mass of $\rm 10.5\times10^{9}\,M_{\odot}$ (Jiang et al. 2007). 
The beam size of the [C II] line observation is 
$\rm 0.66''\times0.45''$ and the deconvolved source size is estimated to be 
$\rm (0.61''\pm0.11'')\times(0.33''\pm0.22'')$ oriented at $\rm PA=65^{\circ}\pm48^{\circ}$. 
No gradient is seen in the line intensity-weighted velocity map of this object. 
The 287 GHz continuum flux density is 
$\rm 3.12\pm0.09\,mJy$ and the source is unresolved in the continuum. 

We calculate the [C II] line luminosity of each source as 
$\rm L_{[CII]}/L_{\odot}=1.04\times10^{-3}S\Delta v\nu_{0}(1+z)^{-1}{D_{L}}^{2}$ 
\citep{solomon05}, where $\rm \nu_{0}=1900.5369\,GHz$ is the rest frame 
[C II] line frequency, $\rm S\Delta v$ is the integrated line flux in 
$\rm Jy\,km\,s^{-1}$, and $\rm D_{L}$ is the luminosity distance in Mpc. 
The derived line luminosities are in the range of 1.6 to 
8.7$\rm \times10^{9}\,L_{\odot}$. 
We also calculate the CO (6-5) line luminosity for J2310+1855, which is 
$\rm (5.4\pm0.5)\times10^{8}\,L_{\odot}$ or $\rm (5.1\pm0.4)\times10^{10}\,K\,km\,s^{-1}\,pc^{2}$. 

We then estimate their FIR luminosities ($\rm L_{FIR}$) by fitting a modified 
black body (i.e., $\rm S_{\nu}\sim {\nu}^{3+\beta}/(exp(h\nu /kT_{dust})-1)$, 
\citealp{priddey01,debreuck03,kovacs06})
to the ALMA continuum flux densities, MAMBO \citep{petric03,wang07,wang08,wang11a},
and available PdBI and SCUBA data \citep{priddey03,priddey08,wang10,wang11a}.
For four of the five objects (except J2054-0005), 
we adopt a dust temperature of $\rm T_{dust}=47$ K and emissivity index of $\rm \beta=1.6$, which are
the mean values found in the sample of high-z FIR luminous quasars \citep{beelen06}. 
This yields the FIR luminosities ($\rm L_{FIR}$) in the 
range 42.5-122.5 $\mu$m of 5$-$17$\rm \times10^{12}\,L_{\odot}$
\footnote{Based on the modified black body model (i.e. $\rm T_{dust}\sim47K$,
$\beta\sim1.6$, \citealp{beelen06}), the commonly used infrared luminosity
integrated from 8 to 1000 $\mu$m is about 1.4 times larger than the 42.5-122.5 $\mu$m
FIR luminosity we used here.}. The estimates of $\rm L_{FIR}$ could be
larger by up to 20\% if a higher dust temperature of 50K is assumed, or lower by $\rm \leq8\%$
if the best fit of $\rm z>4$ quasars from \citet{priddey01}
(i.e. $\rm T_{dust}=41$ K and $\rm \beta=1.95$) is adopted.
J2054-0005 was recently detected by Herschel/SPIRE at 250 $\mu$m 
and 350 $\mu$m, which sample the continuum at wavelengths 
close to or shorter than the peak of the 
starburst-powered thermal dust emission \citep{leipski13}. As the 250 $\mu$m flux  
density might be siginficantly contaminated by the emission 
from the AGN dust torus \citep{leipski13}, we fit $\rm L_{FIR}$ 
for this object with the Herschel 350 $\mu$m, the MAMBO 250 GHz, 
and ALMA 262 GHz measurements, and leave $\rm T_{dust}$ as a free parameter. 
The fit suggests $\rm T_{dust}=52\pm6$ K and $\rm L_{FIR}=(8.0\pm3.3)\times10^{12}\,L_{\odot}$. 
We have listed the derived [C II] and FIR luminosities in Table 3.

\section{Discussion}

\subsection{Distributions of gas, dust, and star formation in the quasar host galaxies}

The new detections of strong [C II] line emission toward the five FIR-luminous 
quasars at z$\sim$6 provide further evidence for active star formation in the 
quasar host galaxies. The [C II] line emission is marginally resolved in all these objects. 
The ALMA observations at $\rm 0.7''$ resolution yield estimates of   
the intrinsic [C II] source sizes (FWHM of the major axis) of 0.3$''$ 
to $\rm 0.6''$ (1.7 to 3.5 kpc). 
Clear velocity gradients have been found in the line intensity-weighted
velocity maps of J2310+1855, J1319+0950, and J0129$-$0035,
which suggests that the gas could be in rotation, and gravitationally
bound, e.g. a non-face-on rotating disk.
The dust continuum flux densities measured in the vicinity of the [C II] line
frequencies for all five objects are consistent with previous MAMBO measurements
at 250 GHz. The continuum sources of four of them (except J1044$-$0125) are  
marginally resolved, indicating deconvolved 
FWHM major axis sizes of 0.2$''$ to 0.4$''$, or 1.2 to 2.3 kpc. 
These results constrain the spatial extent of star forming activity to be 
2.6 to 5.3 kpc in
diameter\footnote{We adopt a source size of 1.5$\rm \times$ the FWHM major axis from
the [C II] intensity map (i.e., full width at 20\% of the peak intensity
for a Gaussian profile). } in the nuclear region. 
We will observe these sources with ALMA in Cycle 1 at 0.2$''$ resolution
to fully resolve the line and dust continuum sources, and measure what fraction of the dust
continuum emission is from the central compact AGN. These observations will finally measure the
star formation rates (SFR) and SFR surface densities in these earliest quasar-starburst systems. 
The higher resolution imaging will also address if the gas components in the nuclear starburst 
region is uncoalesced and show multiple-peak morphology in line emission \citep{walter04}, 
which was suggested by the galaxy merger models of quasar-galaxy 
formation (e.g. \citealp{narayanan08})

In Figure 4, we plot the [C II]-to-FIR luminosity ratios of 
the [C II]-detected z$\geq$5.8 quasars, including the five new 
detections in this work, J1148+5251 \citep{maiolino05}, 
CFHQS J0210$-$0456 at z=6.43 \citep{willott13}, and ULAS J1120+0640 at z=7.08 \citep{venemans12}.
We compare them to samples of [C II]-detected
local normal star forming galaxies, ULIRGs, submillimeter galaxies, and FIR-luminous
quasars at high redshift \citep{malhotra01,luhman03,stacey10,maiolino09,
ivison10,de11,swinbank12,wagg12,pety04,gallerani12,carilli13,valtchanov11,riechers13,marsden05}.
The five z$\sim$6 quasars presented in this work, as well as J1148+5251, show luminosity ratios
of 2.9 to $\rm 5.1\times10^{-4}$, which are comparable to the typical values found in
local ULIRGs and  $\rm 1\leq z\leq 5$ [C II]-detected quasars, and a few to 10 times lower than
that of the disk star forming galaxies and submillimeter galaxies. 
We also notice that the other two z$>$6 quasars with moderate 
FIR luminosities ($\rm 10^{11}$ to $\rm 10^{12}\,L_{\odot}$) 
show higher $\rm L_{[CII]}/L_{FIR}$ than most FIR luminous objects \citep{venemans12,willott13}.  
We estimate the CO (1-0) line luminosities for the five objects from the 
PdBI CO (6-5) detections (last Column of Table 1), assuming a CO excitation 
ladder similar to J1148+5251 \citep{riechers09}.
The calculated [C II]-to-CO (1-0) line luminosity ratios are about 
2400 to 4700, which is slightly higher than the values found in 
local ULIRGs \citep{luhman03}, and the highest value is close to the 
median luminosity ratios found in starburst galaxies (i.e. $\rm L_{[C II]}/L_{CO}\sim4400$, 
\citealp{stacey91,stacey10,swinbank12}). 

The [C II], FIR, and CO luminosity ratios of the five [C II]-detected z$\sim$6 
quasars can be reproduced by PDR 
models \citep{kaufman99,stacey10,luhman03} with gas density 
on orders of $\rm 10^{4}$ to $\rm 10^{5}\,cm^{-3}$ and FUV (6 eV$\rm< h \nu <13.6$ eV) 
radiation field $\rm G_{0}$ of a few $\rm 10^{3}$ to $\rm 10^{4}$ 
(in units of the Habing Field, $\rm 1.6\times10^{-3}\,ergs\,cm^{-2}\,s^{-1}$, \citealp{kaufman99}). 
However, such strong radiation fields will also produce strong [O I] 63$\mu$m line 
emission with intensities comparable or higher than the [C II] line 
\citep{kaufman99,luhman03,carilli13}. Thus future observations 
of the [O I] and other fine structure lines from these objects with 
the full configuration of ALMA will be a crucial test of the 
physical conditions inferred from these PDR models. 
It is also possible that the central AGN has substantial 
contribution to the dust heating and FIR emission, which 
results in lower [C II]-to-FIR luminosity ratios in the nuclear region \citep{luhman03,sargsyan12}. 

We compare the [C II] and CO (6-5) line profiles of the five objects
in Figure 5. 
For four of the five sources, the redshifts measured with [C II] and CO (6-5) 
are consistent within the 1$\sigma$ errors; 
there are no large velocity offsets between the gas components 
traced by [C II] and CO (6-5) lines.
The [C II] FWHM line widths of J2310+1855, J2054$-$0005, and J0129$-$0035 are about 60 
to 115 $\rm km\,s^{-1}$ smaller than the CO (6-5) measurements. But these differences 
are within the 1 to 2 $\sigma$ error bars, as the CO (6-5) 
line width uncertainties for these objects are between 60 and 120 $\rm km\,s^{-1}$ (Table 1). 
The other object J1044$-$0125, shows a larger [C II]
redshift with $\rm \Delta z=z_{[CII]}-z_{CO}=0.0023\pm0.0010$ 
(i.e. a velocity difference of $\rm 100\pm44\,km\,s^{-1}$),
and a much broader [C II] line width (Figure 5). This may indicate  
different kinematical properties between the two gas components in this object.
However, it is also possible that a large fraction of the CO line 
emission is undetected and the CO line width is underestimated 
due to the low signal-to-noise ratio of the line spectrum. 
Thus, deep imaging of the CO line emission with better measurements of the CO line 
profile and spatial distribution is needed to address whether 
the [C II]-emitting gas is more centrally concentrated in J2310+1855, J2054$-$0005, 
and J0129$-$0035, and to understand the origin of the narrower CO line in J1044$-$0125. 

\subsection{Dynamical masses traced by the [C II] line emission}

If we assume a rotating disk geometry for the [C II]-emitting 
gas in these FIR-luminous z$\sim$6 quasars\footnote{We exclude J1044$-$0125 in 
the discussion here as the difference between the [C II] and CO (6-5) line spectra 
may indicate more complicated gas dynamics.}, we can estimate the dynamical 
masses within the [C II]-emitting regions for J2310+1855, J1319+0955, J2054$-$0005, and J0129$-$0035 as   
$\rm M_{dyn}/M_{\odot}\approx1.16\times10^5{v_{cir}}^2D$, 
where $D$ is the disk diameter in kpc from the [C II] measurements (2.6 to 5.3 kpc, 
see Section 4.1) and $\rm v_{cir}$ is
the maximum circular velocity of the gas disk in km $\rm s^{-1}$. 
We estimate $\rm v_{cir}$ as $\rm v_{cir}=0.75FWHM_{[CII]}/sin{\it i}$ 
(i.e., half width at 20\% line maximum), where i is the inclination angle 
between the gas disk and the line of sight ($\rm i=0^{\circ}$ for a face-on disk).
The derived $\rm M_{dyn}{sin}^{2}\,i$ are $\rm 0.9-8.5\times10^{10}\,M_{\odot}$, and 
the uncertainties estimated with the measurement errors 
in $\rm FWHM_{[CII]}$ and [C II] source size are about 10\% to 40\%. 
With the assumption of an inclined disk geometry, we can also have a first guess at  
the disk inclination angle from the [C II] minor and 
major axis ratios (R), i.e. $\rm i=cos^{-1}(R)$, though we should keep 
in mind that the source size measurements 
at current spatial resolution ($\rm \sim0.7''$) still have large 
uncertainties (see Section 3). The estimated inclination angles 
are 46$^{\circ}$, 56$^{\circ}$, 24$^{\circ}$, and 56$^{\circ}$ for 
J2310+1855, J1319+0955, J2054$-$0005, and J0129$-$0035, respectively.
This gives inclination angle-corrected dynamical masses 
of $\rm M_{dyn}=1.3\times10^{10}$ to $\rm 1.2\times10^{11}\,M_{\odot}$. 
We list the derived inclination 
angle-corrected dynamical mass $\rm M_{dyn}$ in Table 3.

There are no published SMBH masses for the four objects yet. 
We calculate the $\rm 1\mu m$ to 8 keV quasar bolometric luminosities for the four sources using an isotropic 
bolometric correction of $\rm L_{bol}=4.2L_{1450}$ 
\citep{runnoe12a,runnoe12b}, where $\rm L_{1450}$ is the rest-frame 
1450 $\rm \AA$ luminosity calculated from the 1450 $\rm \AA$ magnitudes in the discovery 
papers (\citealp{fan00,fan06,jiang09,mortlock09}, Fan et al. 2013, in prep.)
We here assume that the dust extinction from quasar host galaxies at 1450 $\rm \AA$ 
is negligible. 
We then derive the SMBH masses ($\rm M_{BH}$) 
from the quasar luminosity assuming Eddington accretion. 
The resulting mass ratios $\rm M_{BH}/M_{dyn}$ lie in the range of 0.012 to 0.030 (see Table 3).
Note that the $\rm M_{BH}$ and $\rm M_{BH}/M_{dyn}$ values could be even larger 
if the typical ratio between quasar bolometric luminosity and Eddington 
luminosity is less than unity (e.g. \citealp{de11}) or the 1450 $\rm \AA$ 
luminosities are obscured by the dust in the quasar host 
galaxies \citep{maiolino04,gallerani10,hjorth13}. 


The [C II]-based $\rm M_{BH}/M_{dyn}$ values are 
consistent with the previous CO estimates of the median SMBH-bulge mass ratio 
of these z$\sim$6 quasars \citep{wang10} and agrees with the results
found with other high-z FIR and CO luminous
quasars that the SMBH-bulge mass ratios are 10 to 30 times higher than
the average value of 0.0014 found in local normal galaxies
\citep{marconi03,walter04,riechers08,wang10,coppin08,venemans12}.
However, one should be cautious with these [C II]-based 
$\rm M_{dyn}$ values, as the detected [C II] line 
emission may trace only the intense star forming region in 
the very centers of quasar host galaxies and not extend as far as the stellar bulge. This  
may underestimate the total masses within the spheroidal stellar bulges. 
Additionally, we notice that there are still large uncertainties in the 
intrinsic source morphology measurements; the deconvolved minor axis and position angle measurements 
of some of the objects show large error bars. This can introduce significant 
uncertainties in the inclination and dynamical mass estimate.
Further ALMA imaging of these FIR-luminous z$\sim$6 quasars at $\rm \lesssim0.2''$ will 
be crucial to better constrain the distribution  
and dynamical properties of the gas components, address how well the atomic/molecular  
line emission traces the dynamical masses 
of the quasar host galaxies, and determine whether the high $\rm M_{BH}/M_{dyn}$ 
ratios is common in the massive quasar-starburst systems in the early universe.

\subsection{Search for [CII] line emitters associated with the quasar environments}

Given the sensitivity of the ALMA Band 6 and Band 7 data, and the large spectral
bandwidth (7.5 GHz, or $\sim$9000 $\rm km\,s^{-1}$), it is plausible 
that serendipitous line emission from companion objects will
be detected in our data. Although the volume is relatively small over the
five fields (each field was observed across four 1.875GHz spectral windows, 
covering a redshift range $\rm \Delta z \sim 0.2$ which, for a 
primary beam with FWHM$\rm \sim 22''$ results in a total surveyed volume of $\sim
280$ Mpc$^{3}$), recent ALMA surveys of submm luminous starburst
galaxies at $z \sim 4.4$ suggest a strong evolution in the [C II]
luminosity function out to these early cosmic times (Swinbank et al.
2012). We use a Bayesian search algorithm developed to efficiently
search for broad, weak line emission in large spectral line data
cubes (Lentati et al. 2012), and use the evidence for candidate detections
to calculate the probability that detected candidates are 'real' relative to being noise.
We do not find any candidate detections of serendipitous line emission
in our data with probabilities above $25\%$. At the current sensitivity the
volume sampled is not sufficient to place constraints on the possible
evolution of the luminosity function out to $\rm z \sim 6$, with source 
counts (i.e., $\rm 10^{-3}\,Mpc^{-3}$, Swinbank et al. 2012) derived from
the [CII] luminosity function predicting $\rm \sim 0.3$ detection above a
3$\sigma$ luminosity limit $\rm L_{[CII]} > 2.1\times 10^8$ L$_{\odot}$. 
However the sensitivity of ALMA at bands 6 and 7 means that deeper 
observations in Cycle 1 of similar volumes will
allow us to place strong constraints on the obscured star-formation
properties of galaxies in the environments of the quasars.


\section{Summary}

We detected [C II] fine structure line and dust continuum emission from the host galaxies 
of five quasars at z$\sim$6, using ALMA in the Cycle 0 phase 
at $\rm \sim0.7''$ resolution. 
Complementary IRAM observations of the CO (6-5) line and 250 GHz dust continuum 
emission from the z=6.00 quasar J2310+1855 are also presented.
Our ALMA observations measure the FWHM major axis sizes of the [C II] emission 
from the five objects to be 0.3$''$ to $\rm 0.6''$ (1.7 to 3.5 kpc) and the sizes of the dust 
continuum source for four of them to 0.2$''$ to $\rm 0.4''$ (1.2 to 2.3 kpc). 
The detections of [C II] line and dust continuum emission indicate active 
star formation in the central few kpc region of the quasar host galaxies.
The derived [C II]-to-FIR luminosity 
ratios are of the order $\rm 10^{-4}$, which are comparable to the typical 
values found in local ULIRGs and other FIR-luminous quasars at high redshift. 
The intensity-weighted velocity maps of J2310+1855, J1319+0950, 
and J0129$-$0035 show velocity gradients. 
Such velocity gradients are consistent with rotation, although they are not uniquely interpreted as such.
We estimate the dynamical masses within the [C II]-emitting region for four of the five objects assuming 
that the gas is distributed in a rotating disk. The derived ratios between the SMBH masses 
and the dynamical masses are one order of magnitude higher than that of
local normal galaxies. 

The detections of [C II] 158$\mu$m line emission from quasar host galaxies at z$\sim$6 
have demonstrated the power of ALMA in observing signatures of star formation at 
the earliest cosmic epoch. 
With the full configuration of ALMA, we 
should be able to resolve the line and dust continuum emission on kpc or sub-kpc scales, 
which will measure the surface densities of the gas components and star 
forming activity in the nuclear region, and better address the gas kinetics and dynamical 
masses of the spheroidal quasar host galaxies. The full frequency coverage of ALMA 
will also allow a search of other ionized/atomic interstellar cooling lines from 
these FIR-luminous z$\sim$6 quasars to measure the physical conditions 
(e.g. density, rediation field, temperature, etc.)  
of the interstellar medium in these earliest quasar-starburst systems. 



\acknowledgments
We thank M. Lacy at the National Radio Astronomy Observatory 
for help with the observation and data analysis. 
This work is based on observations carried out with ALMA (NRAO), the Max Planck Millimeter
Bolometer Array (MAMBO) on the IRAM 30m telescope, and the Plateau
de Bure Interferometer. The National
Radio Astronomy Observatory (NRAO) is a facility of the National
Science Foundation operated under cooperative agreement by Associated
Universities, Inc. This paper makes use of the following ALMA data:
ADS/JAO.ALMA\# 2011.0.00206.S . ALMA is a partnership
of ESO (representing its member states), NSF (USA) and NINS (Japan),
together with NRC (Canada) and NSC and ASIAA (Taiwan), in
cooperation with the Republic of Chile. The Joint ALMA Observatory
is operated by ESO, AUI/NRAO and NAOJ. IRAM is supported
by INSU/CNRS (France), MPG (Germany) and IGN (Spain).
Frank Bertoldi and Fabian Walter acknowledge 
support through the DFG priority program 1573 and the SFB 956.
Desika Narayanan acknowledges support from the NSF via grant AST-1009452. 
Kirsten Knudsen acknowledges support from the Swedish Research Council.
X. Fan acknowledges support from NSF grant AST 08-06861 and 11-07682 and 
a David and Lucile Packard Fellowship.



{\it Facilities:} \facility{ALMA}, \facility{IRAM: 30m (MAMBO)}, \facility{IRAM: Interferometer Europe}

\clearpage

\begin{table}
{\scriptsize \caption{Summary of previous observations}
\begin{tabular}{lcccccc}
\hline \noalign{\smallskip}
Source & $\rm m_{1450}$ & $\rm S_{250}$  & $\rm z_{CO}$ & $\rm FWHM_{CO}$ & $\rm S\Delta v_{CO(6-5)}$ 
& $\rm L_{CO(1-0)}$ \\
       &     & mJy &             & $\rm km\,s^{-1}$ & $\rm Jy\,km\,s^{-1}$ & $\rm 10^{5}\,L_{\odot}$\\
 (1)   & (2) & (3) & (4) & (5) & (6) & (7) \\
\noalign{\smallskip} \hline \noalign{\smallskip}
SDSS J231038.88+185519.7 &19.30 & 8.29$\pm$0.63 & 6.0025$\pm$0.0007 & 456$\pm$64 & $\rm 1.52\pm0.13$ & 32.1 \\
ULAS J131911.29+095051.4   &19.65 & 4.20$\pm$0.65 & 6.1321$\pm$0.0012 &537$\pm$123 & $\rm 0.43\pm0.09$ & 9.4  \\
SDSS J205406.49$-$000514.8 &20.60 & 2.38$\pm$0.53 & 6.0379$\pm$0.0022 & 360$\pm$110 &$\rm 0.34\pm0.07$ & 7.3 \\
SDSS J012958.51$-$003539.7 &22.28 & 2.37$\pm$0.49 & 5.7794$\pm$0.0008 & 283$\pm$87 & $\rm 0.37\pm0.07$ & 7.4 \\
SDSS J104433.04$-$012502.2 &19.21 & 1.82$\pm$0.43 & 5.7824$\pm$0.0007 & 160$\pm$60 & $\rm 0.21\pm0.04$ & 4.2 \\
\noalign{\smallskip} \hline
\end{tabular}\\
Note -- Column (1), name; Column (2), magnitudes at rest-frame 1450 $\rm \AA$ 
(Fan et al. 2013, in prep; \citealp{mortlock09,jiang08,jiang09,fan00}); 
Column (3), continuum flux density at 250 GHz; Column (4) and (5), 
redshift and FWHM line width of the CO (6-5) line; Column (6) CO (6-5) line flux of 
the five objects; Column (7) 
CO (1-0) line luminosity calculated from the CO (6-5) line emission assuming
a CO excitation ladder similar to J1148+5251 \citep{riechers09}.
The CO (6-5) line measurements for J2310+1855 are from this work, and from \citet{wang10,wang11a} 
for the other four sources.
}
\end{table}
\begin{table}
{\scriptsize \caption{ALMA observations}
\begin{tabular}{lcccccccc}
\hline \noalign{\smallskip}
Source & $\rm t_{on}$ & $\rm z_{[CII]}$ & S$\Delta$v & $\rm FWHM_{[CII]}$ & $\rm rms_{[C II]}$ & $\rm {\nu}_{con}$ & $\rm S_{con}$ & $\rm rms_{con}$ \\
       & min & & $\rm Jy\,km\,s^{-1}$ & $\rm km\,s^{-1}$ & $\rm mJy\,Beam^{-1}$ & GHz & mJy & $\rm mJy\,Beam^{-1}$ \\
 (1)   & (2) & (3) & (4) & (5) & (6) & (7) & (8) & (9)  \\
\noalign{\smallskip} \hline \noalign{\smallskip}
J2310+1855  &50 & 6.0031$\pm$0.0002 & 8.83$\pm$0.44 & 393$\pm$21& 0.5 & 263 & 8.91$\pm$0.08 &0.06  \\
J1319+0950  &80 & 6.1330$\pm$0.0007 & 4.34$\pm$0.60 & 515$\pm$81& 0.7 & 258 & 5.23$\pm$0.10 & 0.08 \\
J2054$-$0005&40 & 6.0391$\pm$0.0001 & 3.37$\pm$0.12 & 243$\pm$10& 0.4 & 262 & 2.98$\pm$0.05 &0.04   \\
J0129$-$0035&60 & 5.7787$\pm$0.0001 & 1.99$\pm$0.12 & 194$\pm$12& 0.4 & 287 &2.57$\pm$0.06 &0.05  \\ 
J1044$-$0125&87 & 5.7847$\pm$0.0007 & 1.70$\pm$0.30 & 420$\pm$80& 0.6 & 287 & 3.12$\pm$0.09 &0.09  \\
\noalign{\smallskip} \hline
\end{tabular}\\
Note -- Column (1), source; Column (2), on-source time of the ALMA [C II] observations; Column (3), (4), and (5), redshift, flux, and FWHM line width 
of the [C II] line fitted to a single Gaussian line profile; 
Column (6), ALMA 1$\sigma$ line sensitivity 
bin to a channel width of 62.5 MHz ($\rm \sim70\,km\,s^{-1}$); 
Column (7) and (8),
frequency and flux density of the continuum measured with the line-free windows;
Column (9), ALMA 
1$\sigma$ rms continuum sensitivity integrated over a total bandwidth of $\rm 5.8-6\,GHz$.
The flux errors quoted here are from our fitting process described in Section 3. 
The calibration uncertainties are not included here.
}
\end{table}
\begin{table}
{\scriptsize \caption{Luminosities and Dynamical masses}
\begin{tabular}{lccccccc}
\hline \noalign{\smallskip}
Source & $\rm L_{[CII]}$ & $\rm L_{FIR}$ & $\rm L_{bol}$ & $\rm M_{BH}$ & $\rm M_{dyn}{sin}^2i$ & $\rm M_{dyn}$ & $\rm M_{BH}/M_{dyn}$ \\
      & $\rm 10^{9}\,L_{\odot}$ & $\rm 10^{12}\,L_{\odot}$ &$\rm 10^{13}\,L_{\odot}$ &$\rm 10^{9}\,M_{\odot}$ & $\rm 10^{10}\,M_{\odot}$ & $\rm 10^{10}\,M_{\odot}$ & \\
 (1)   & (2) & (3) & (4) & (5) & (6) & (7) &(8)  \\
\noalign{\smallskip} \hline \noalign{\smallskip}
J2310+1855   & 8.7$\pm$1.4 &17.0$\pm$1.8 &9.3 & 2.8 & 4.9$\pm$0.6 & 9.6 & 0.030 \\
J1319+0950   & 4.4$\pm$0.9 &10.7$\pm$1.3 &7.0 & 2.1 & 8.5$\pm$2.9 &12.5   & 0.017 \\
J2054$-$0005 & 3.3$\pm$0.5 &8.0$\pm$3.3  &2.8 &0.86 &  1.2$\pm$0.2 & 7.2 & 0.012 \\
J0129$-$0035 & 1.8$\pm$0.3 &4.6$\pm$0.6 &0.57  & 0.17 &  0.9$\pm$0.2 & 1.3 & 0.013 \\
J1044$-$0125 & 1.6$\pm$0.4 &5.5$\pm$0.7 &11.6  & 10.5 & -- & -- & -- \\ 
\noalign{\smallskip} \hline
\end{tabular}\\
Note -- Column (1), source; Column (2), [C II] line luminosity; Column (3), 
FIR luminosity in the wavelength range from 42.5$\mu$m to 122.5$\mu$m (see Section 3 for details). 
We considered both measurement errors listed in Table 2 and the 15\% to 20\% 
calibration uncertainties in the calculation of $\rm L_{[CII]}$ and $\rm L_{FIR}$. 
Column (4), quasar bolometric luminosities. The bolometric luminosity 
for J1044-0125 is taken from \citet{jiang06}, and we estimate the bolometric luminosities 
for the other four objects from their rest-frame 
1450 $\rm \AA$ magnitude, assuming an isotropic
bolometric correction of $\rm L_{bol}=4.2L_{1450}$ \citep{runnoe12a,runnoe12b}; 
Column (5), SMBH masses. The SMBH mass 
of J1044$-$0125 is calculated from the quasar C IV line emission (Jiang et al. 2007), and
we estimate the SMBH masses for the other four sources from the bolometric luminosities 
assuming Eddington accretion; Column (6), dynamical mass without 
inclination angle correction, estimated from the [C II] line width and source 
size (see Section 4.2); Column (7), inclination angle-corrected dynamical mass 
within the [C II]-emitting region. We estimate the disk inclination angle from the [C II] minor
and major axis ratio as $\rm i=46^{\circ}$, $\rm 56^{\circ}$, $\rm 24^{\circ}$, 
and $\rm 56^{\circ}$ for J2310+1855, J1319+0950, J2054$-$0005, 
and J0129$-$0035, respectively;  Column (8), SMBH-bulge mass ratio. 
}
\end{table}

\begin{figure}[h]
\includegraphics[height=1.7in]{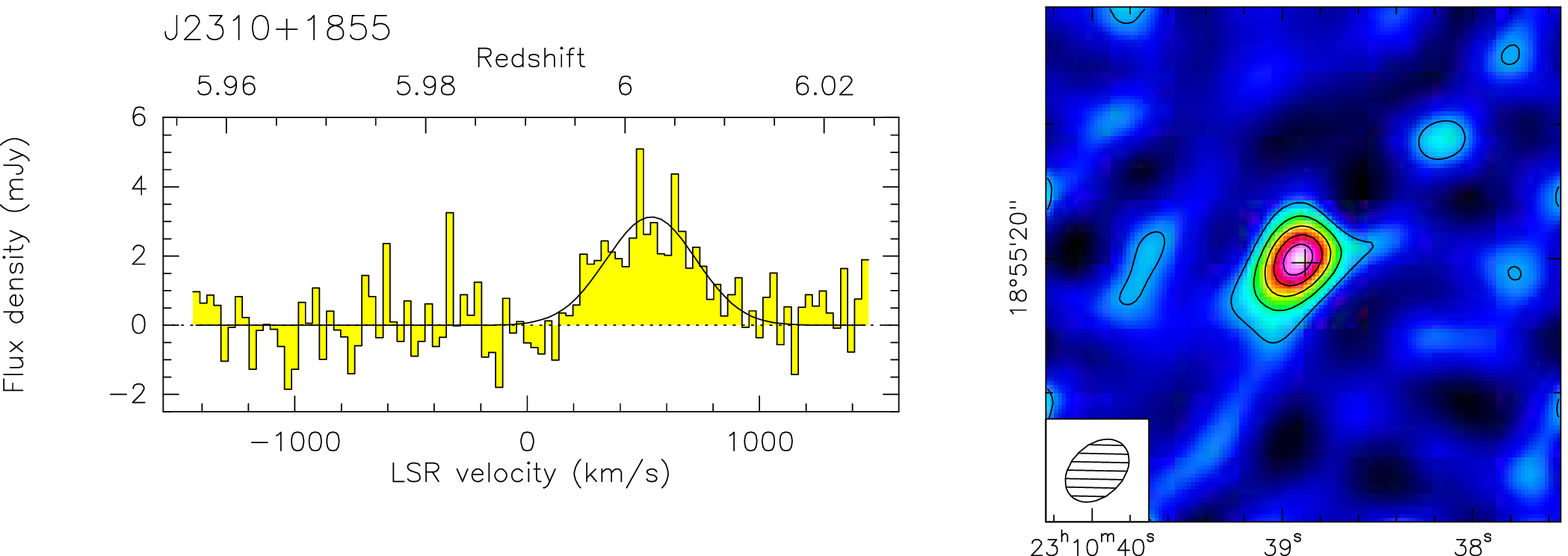}
\vskip -1.7in
\hspace*{5.0in}
\includegraphics[height=1.7in]{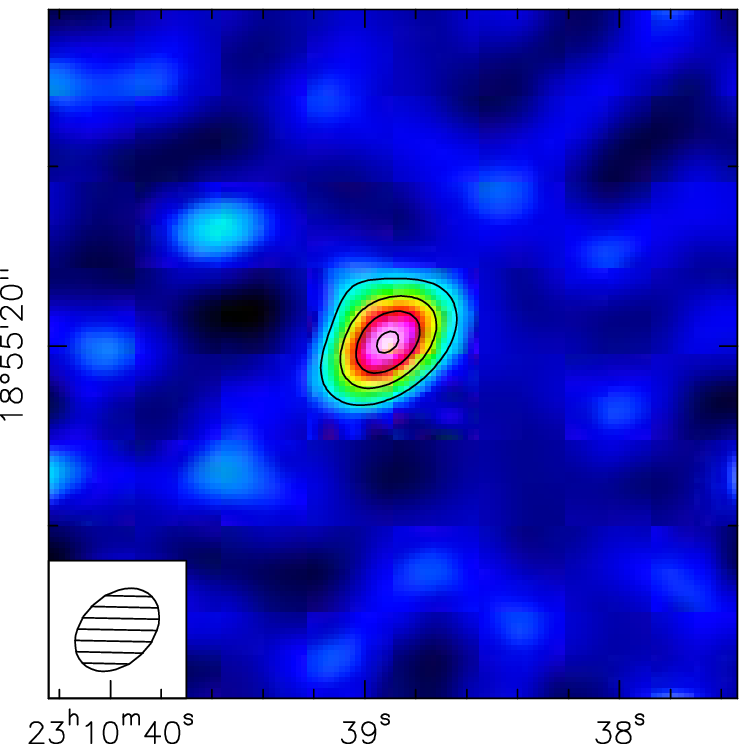}
\caption{PdBI observation of CO (6-5) line and dust continuum emission at 99GHz 
from J2310+1855. The left panel shows the CO (6-5) line spectrum 
binned to 30 $\rm km\,s^{-1}$ channels. 
The solid line is a Gaussian fit to the line spectrum. The middle 
panel shows the intensity map of the CO (6-5) line emission. 
The 1$\sigma$ rms noise of the map is $\rm 0.13\,Jy\,km\,s^{-1}\,Beam^{-1}$ and the  
contours in steps of 2$\sigma$. 
The beam size of $\rm 5.4''\times3.9''$ is plotted on the bottom 
left. The cross denotes the position of the optical quasar. The right panel 
is the continuum map at 99 GHz with 1$\sigma$ rms of $\rm 0.05\,mJy\,Beam^{-1}$, 
and the contours in steps of 2$\sigma$.}
\end{figure}

\begin{figure}[h]
\includegraphics[height=1.75in]{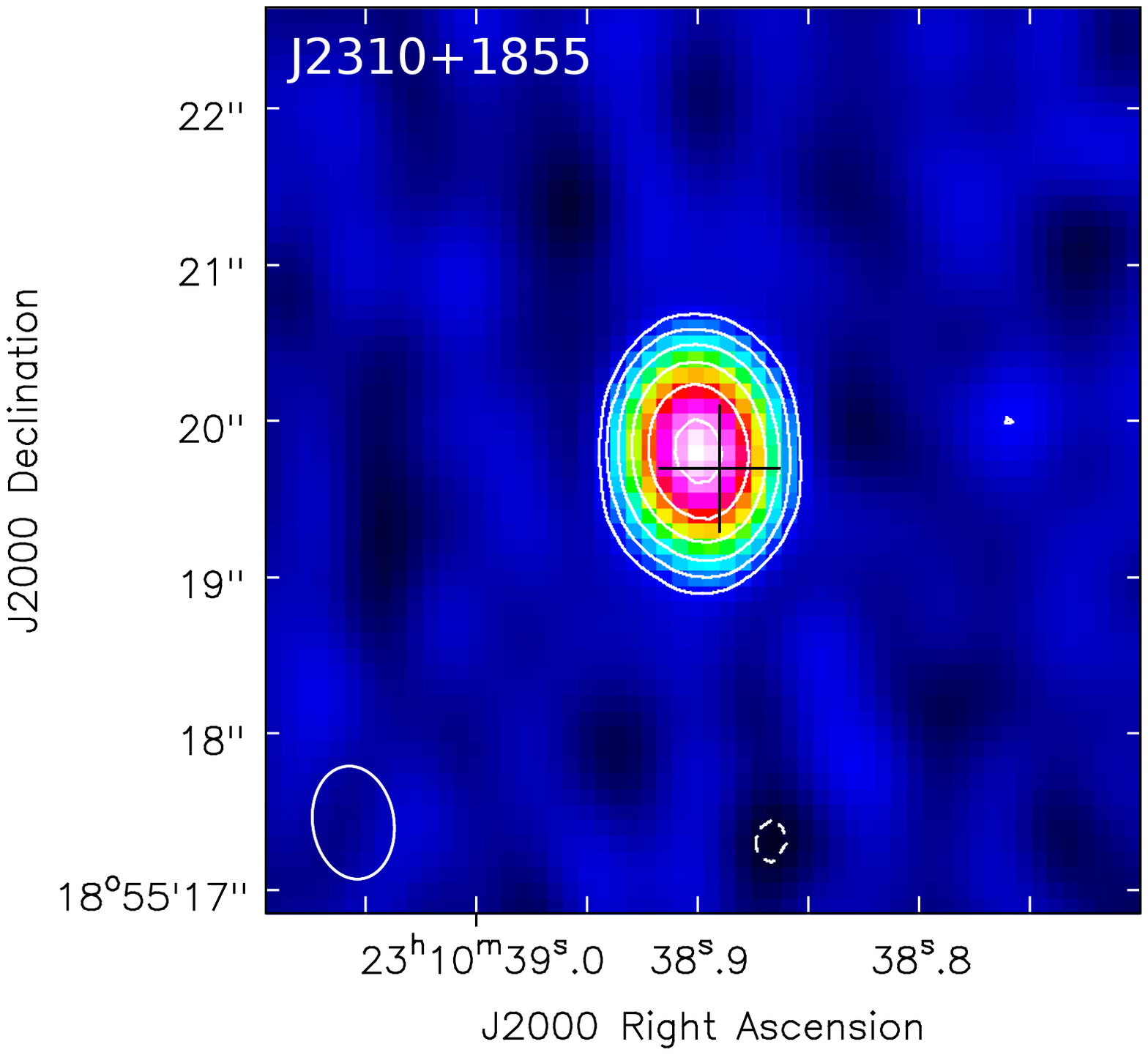}
\vskip -1.75in
\hspace*{2.0in}
\includegraphics[height=1.75in]{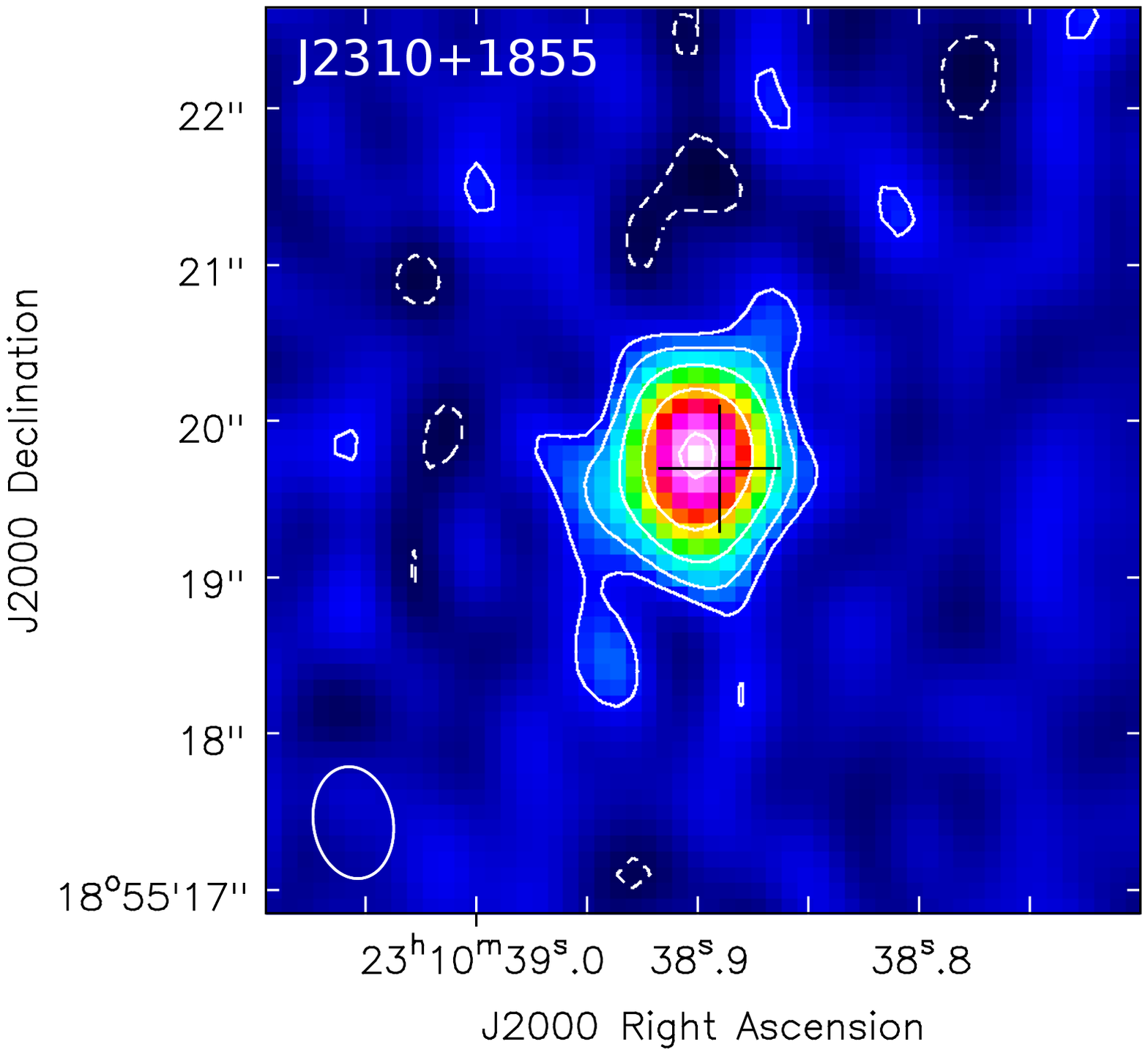}
\vskip -1.75in
\hspace*{4.0in}
\includegraphics[height=1.75in]{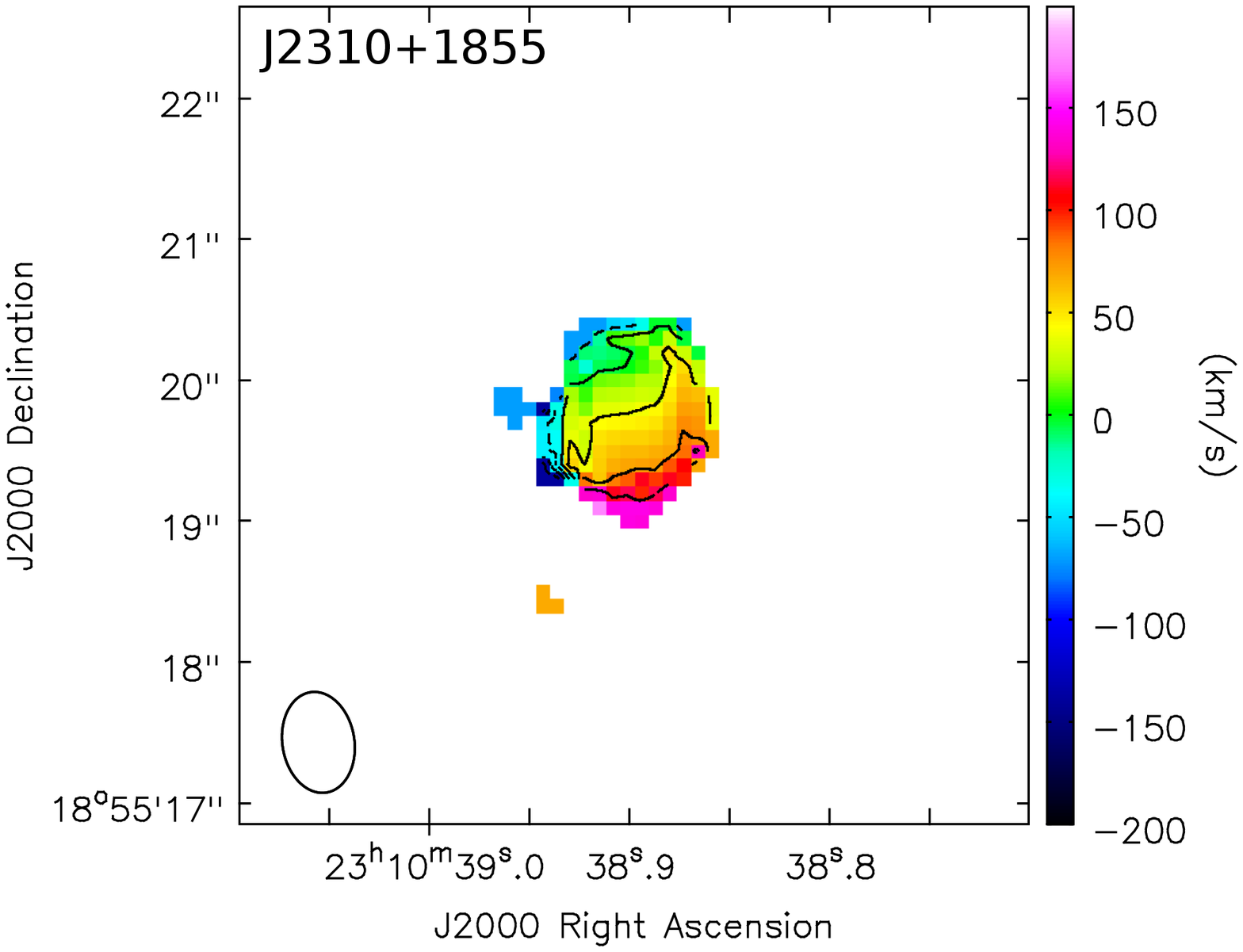}\\
\includegraphics[height=1.8in]{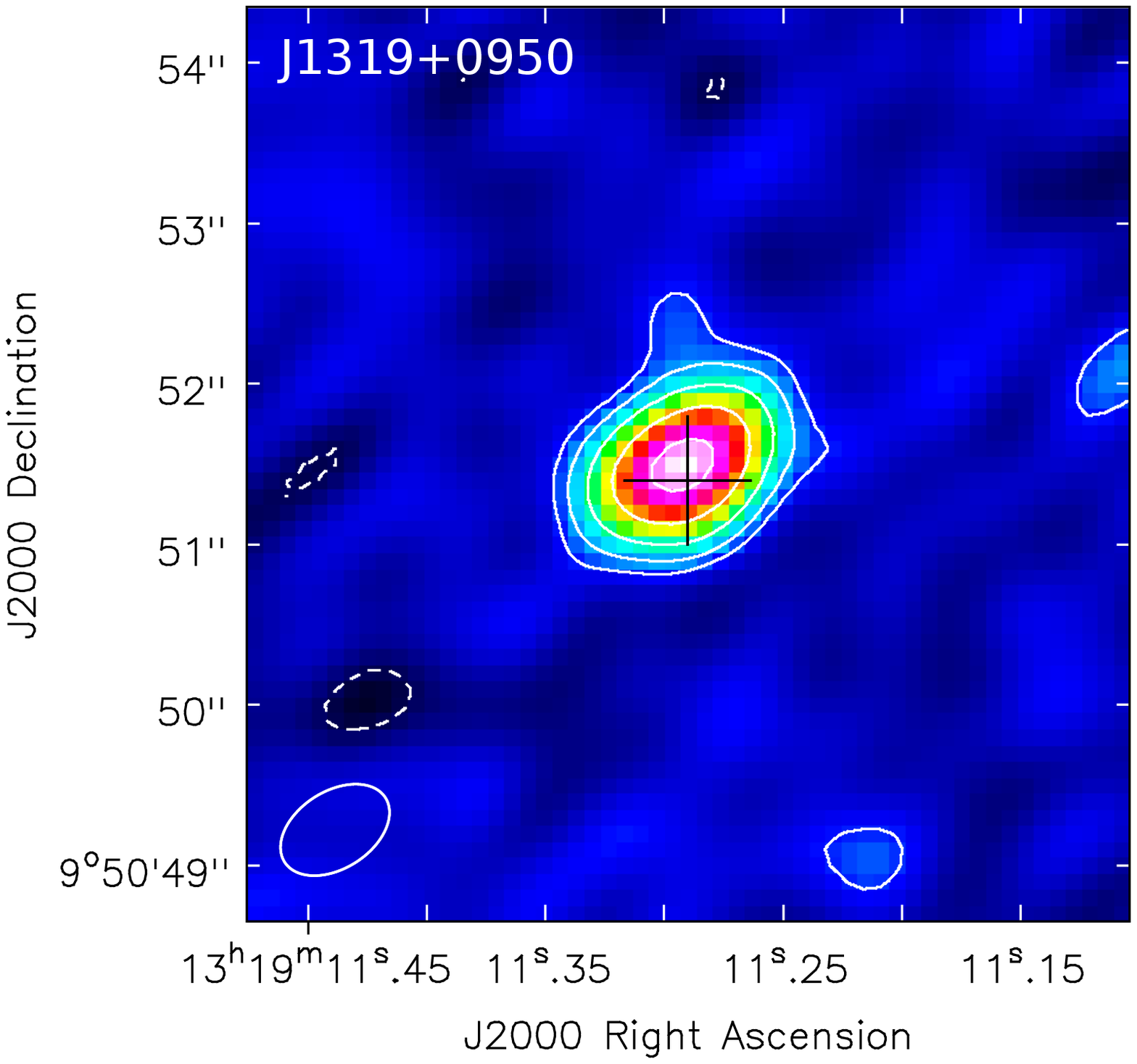}
\vskip -1.8in
\hspace*{2.0in}
\includegraphics[height=1.8in]{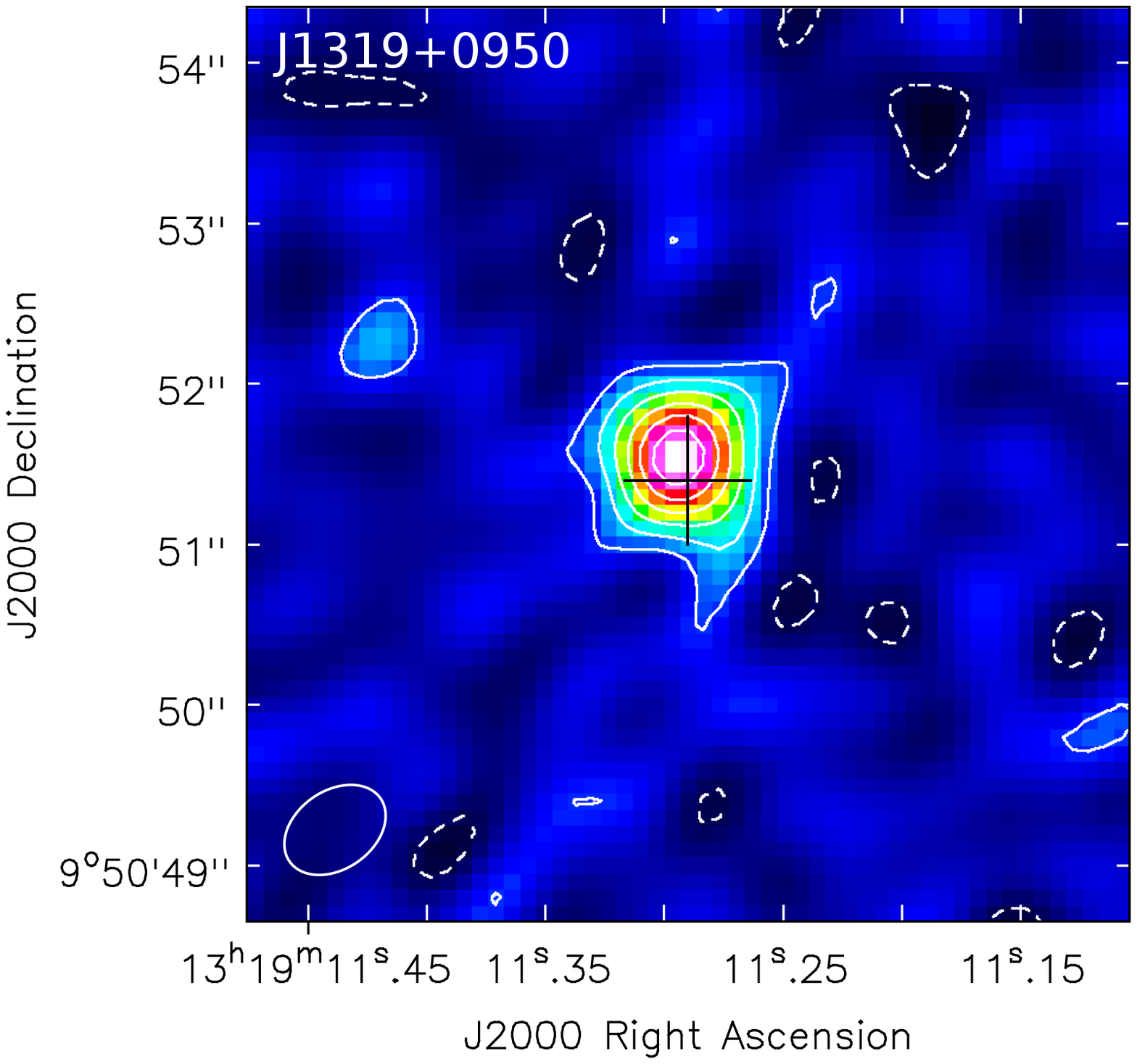}
\vskip -1.8in
\hspace*{4.0in}
\includegraphics[height=1.8in]{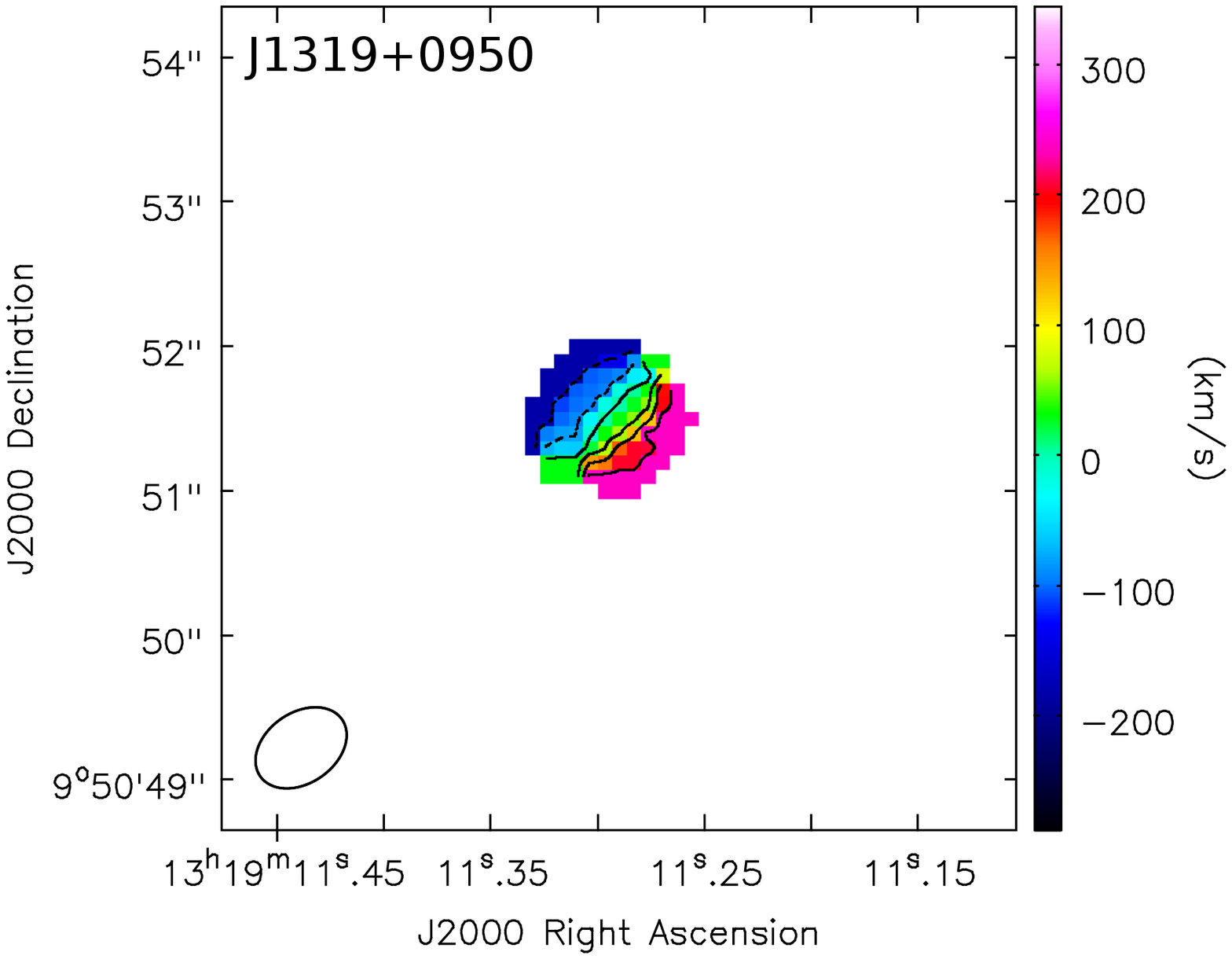}\\
\includegraphics[height=1.7in]{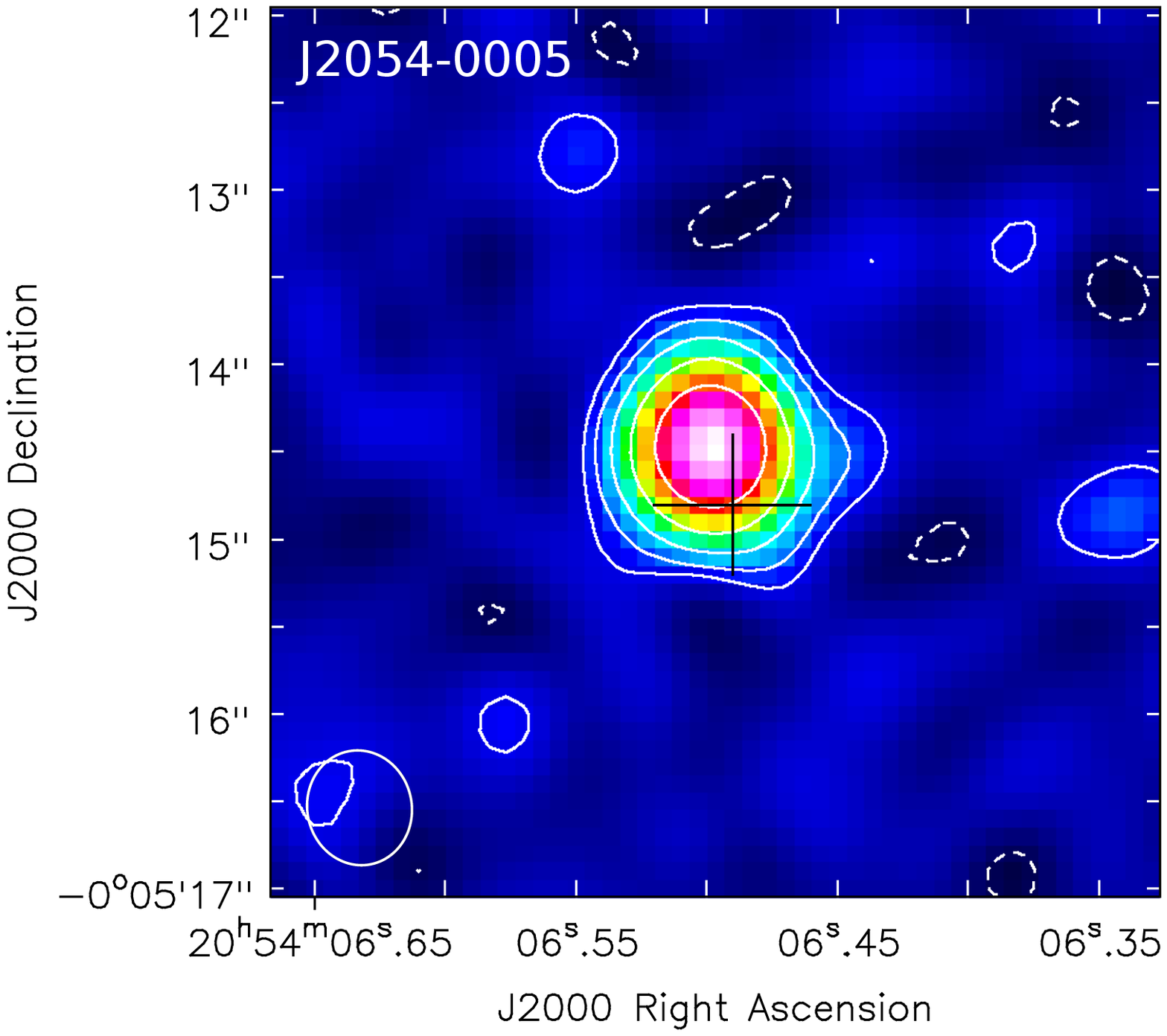}
\vskip -1.7in
\hspace*{2.0in}
\includegraphics[height=1.7in]{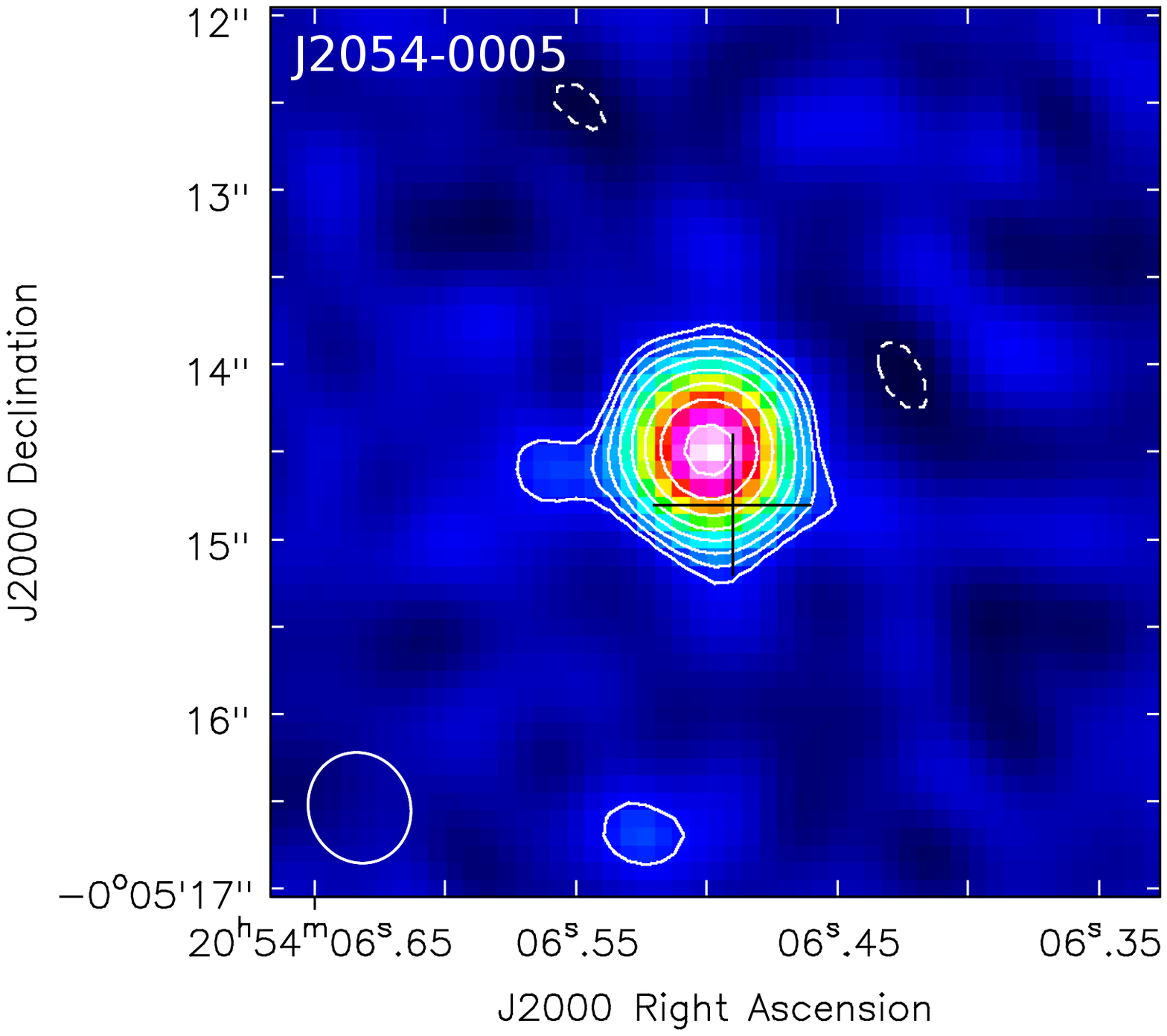}
\vskip -1.7in
\hspace*{4.0in}
\includegraphics[height=1.7in]{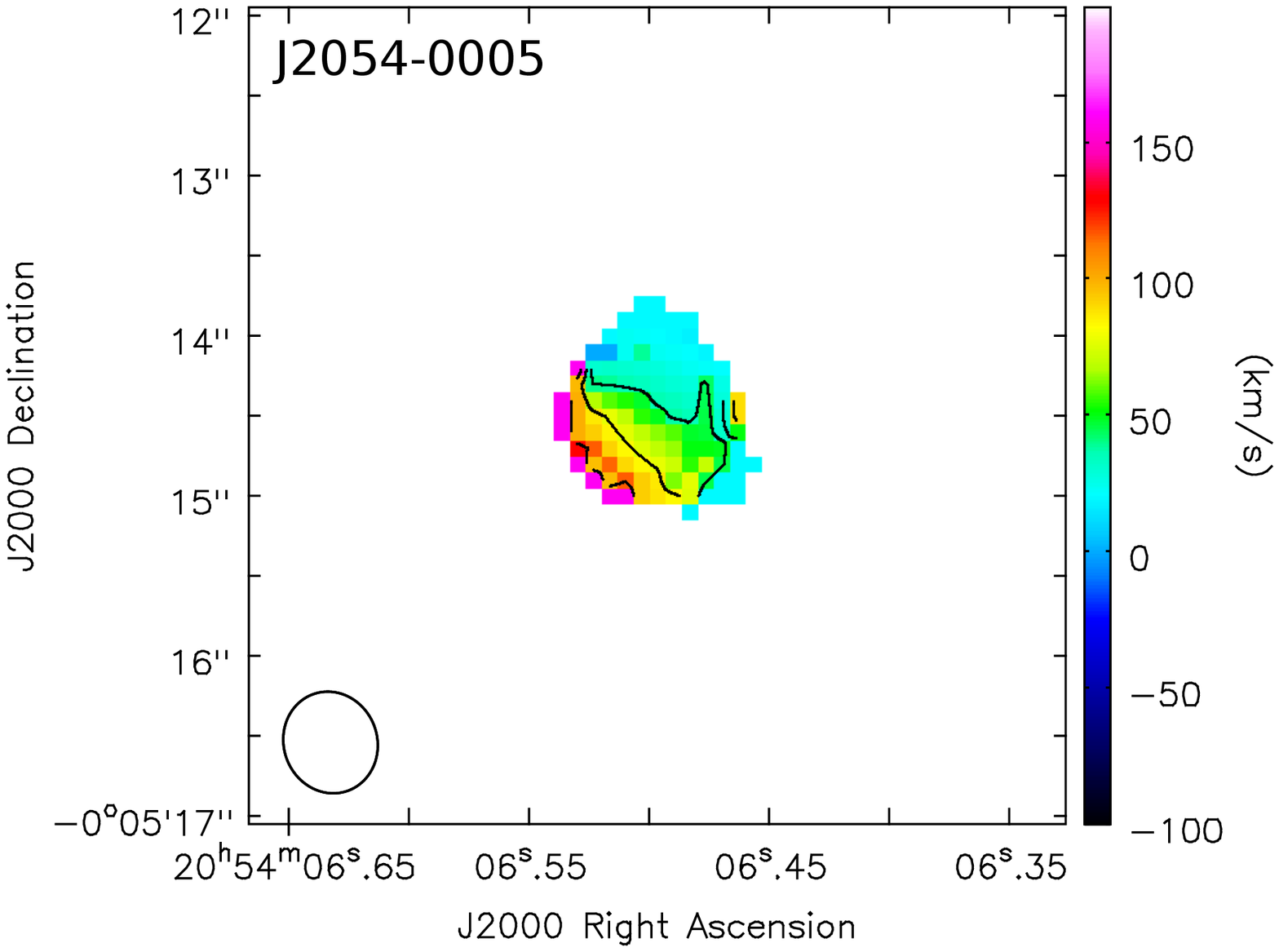}\\
\caption{\scriptsize The dust continuum map (left), [C II] line velocity-integrated map (middle), and line velocity
maps (right) of the five new [C II] detections. 
We calculate the line intensity-weighted velocity 
map using pixels detected at $\rm \geq4\sigma$ in each case. 
The black crosses show the position of the 
optical quasar from the discovery paper (Fan et al. 2013, in prep., 
\citealp{mortlock09,jiang08,jiang09,fan00}). The sizes of the synthesized beams are plotted 
in the bottom-left of each panel. 
{\bf J2310+1855} -- The continuum contours are [-2, 2, 4, 8, 32, 64]$\rm \times0.1\,mJy\,Beam^{-1}$, and the line 
contours are [-2, 2, 4, 8, 16, 32]$\rm \times0.15\,Jy\,Beam^{-1}\,km\,s^{-1}$. 
The velocity contours are [-1, 0, 1, 2, 3]$\rm \times40\,km\,s^{-1}$. 
The 1$\sigma$ rms noise is $\rm 0.06\,mJy\,Beam^{-1}$ 
for the continuum map and $\rm 0.14\,Jy\,km\,s^{-1}\,Beam^{-1}$ for the line map.
 {\bf J1319+0950} -- The contours are 
[-2, 2, 4, 8, 16, 32]$\rm \times0.1\,mJy\,Beam^{-1}$ for the continuum, 
[-2, 2, 4, 6, 8, 10, 12]$\rm \times0.18\,Jy\,Beam^{-1}\,km\,s^{-1}$ for the line, and 
[-2, -1, 0, 1, 2, 3]$\times75\,km\,s^{-1}$ for the velocity map. 
The 1$\sigma$ rms noise values are $\rm 0.08\,mJy\,Beam^{-1}$ 
and $\rm 0.18\,Jy\,km\,s^{-1}\,Beam^{-1}$ for the continuum and line maps, respectively.
{\bf J2054$-$0005} -- The continuum contours are
[-2, 2, 4, 8, 16, 32, 64]$\rm \times0.04\,mJy\,Beam^{-1}$, the 
line contours are [-2, 2, 2.83, 4, 5.66, 8, 11.31, 16, 
22.63]$\rm \times0.10\,Jy\,Beam^{-1}\,km\,s^{-1}$, and the contours in the velocity map are 
[0, 1, 2]$\rm \times40\,km\,s^{-1}$. 
The 1$\sigma$ rms noise is $\rm 0.04\,mJy\,Beam^{-1}$ for the continuum map 
and $\rm 0.08\,Jy\,km\,s^{-1}\,Beam^{-1}$ for the line map.
{\bf J0129$-$0035} -- The contours are
[-2, 2, 4, 8, 16, 32]$\rm \times0.1\,mJy\,Beam^{-1}$ for the continuum, 
[-2, 2, 4, 8, 16, 32]$\rm \times0.075\,Jy\,Beam^{-1}\,km\,s^{-1}$ for the line, and 
[-3, -2, -1, 0]$\rm \times20\,km\,s^{-1}$ for the velocity map. 
The 1$\sigma$ rms noise is $\rm 0.05\,mJy\,Beam^{-1}$ for the continuum map 
and $\rm 0.08\,Jy\,km\,s^{-1}\,Beam^{-1}$ for the line map.
{\bf J1044$-$0125} -- The contours are
[-2, 2, 4, 8, 16]$\rm \times0.08\,mJy\,Beam^{-1}$ for the continuum and 
[-2, 2, 4, 6, 8]$\rm \times0.14\,Jy\,Beam^{-1}\,km\,s^{-1}$ for the line. 
The 1$\sigma$ rms noise is $\rm 0.09\,mJy\,Beam^{-1}$ for the continuum map 
and $\rm 0.14\,Jy\,km\,s^{-1}\,Beam^{-1}$ for the line map. 
The zero velocity in the velocity maps correspond  
to the CO redshifts listed in Table 1.}
\end{figure}
\begin{figure}[h]
\figurenum{2}
\includegraphics[height=1.7in]{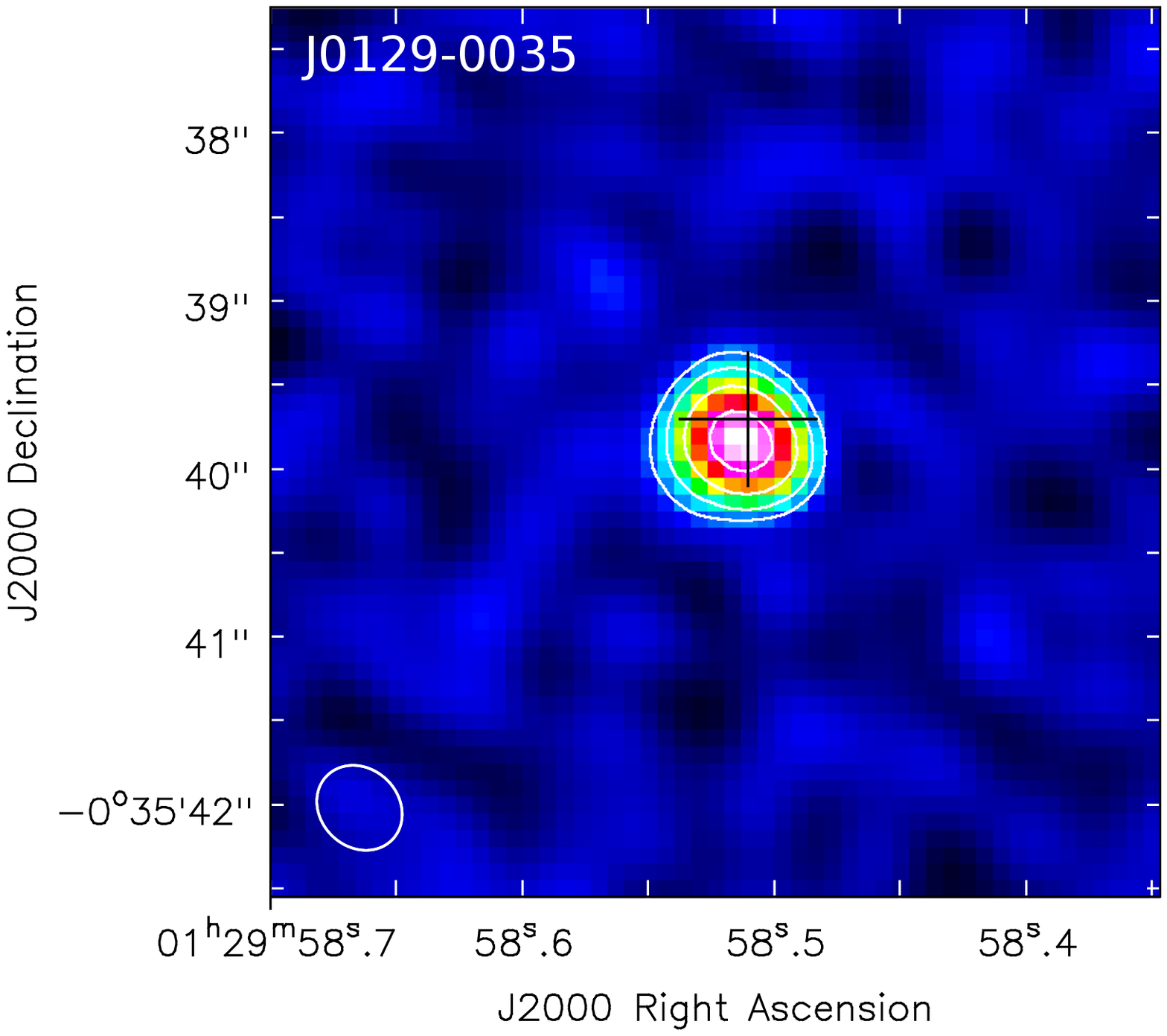}
\vskip -1.7in
\hspace*{2.0in}
\includegraphics[height=1.7in]{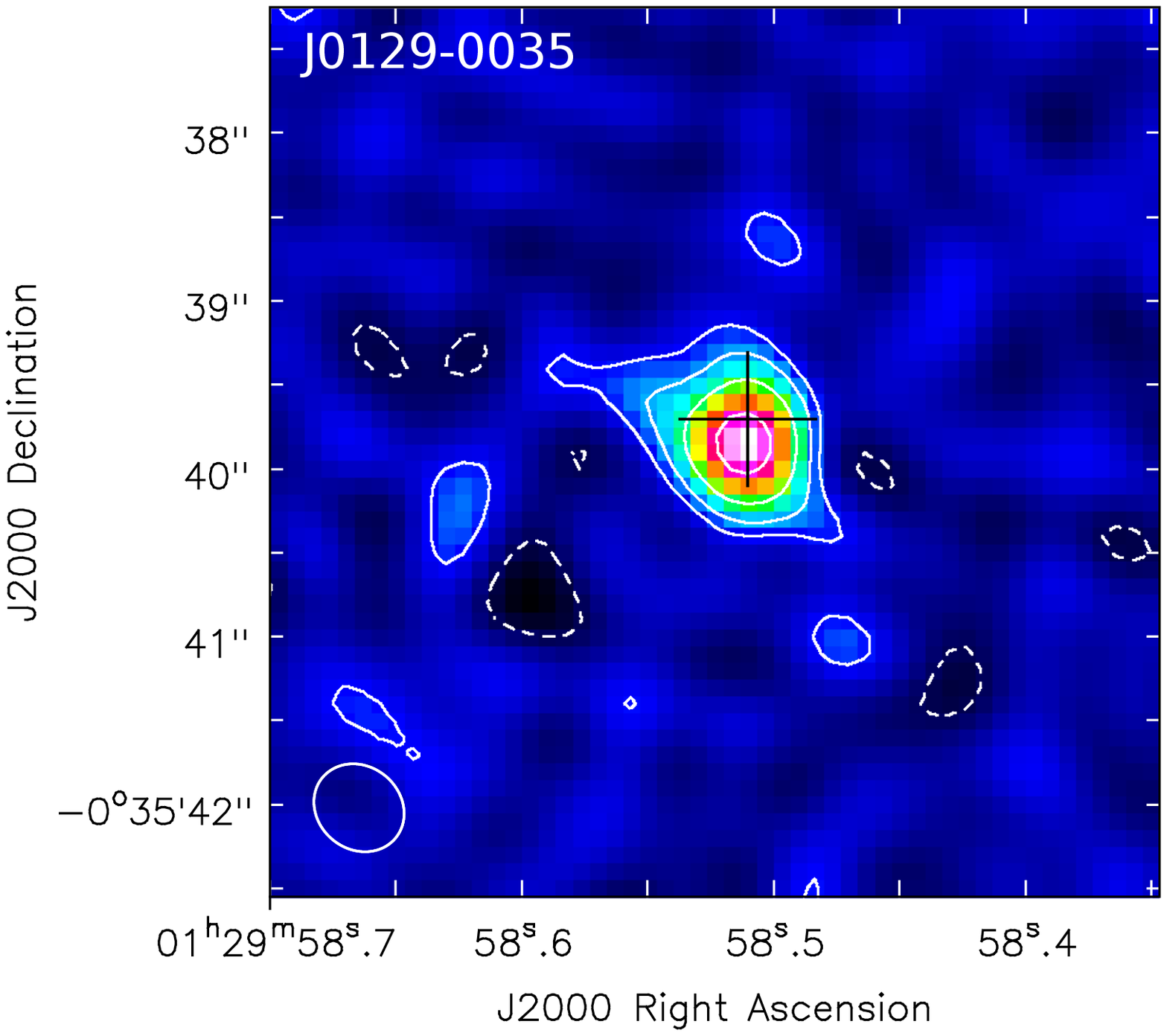}
\vskip -1.7in
\hspace*{4.0in}
\includegraphics[height=1.7in]{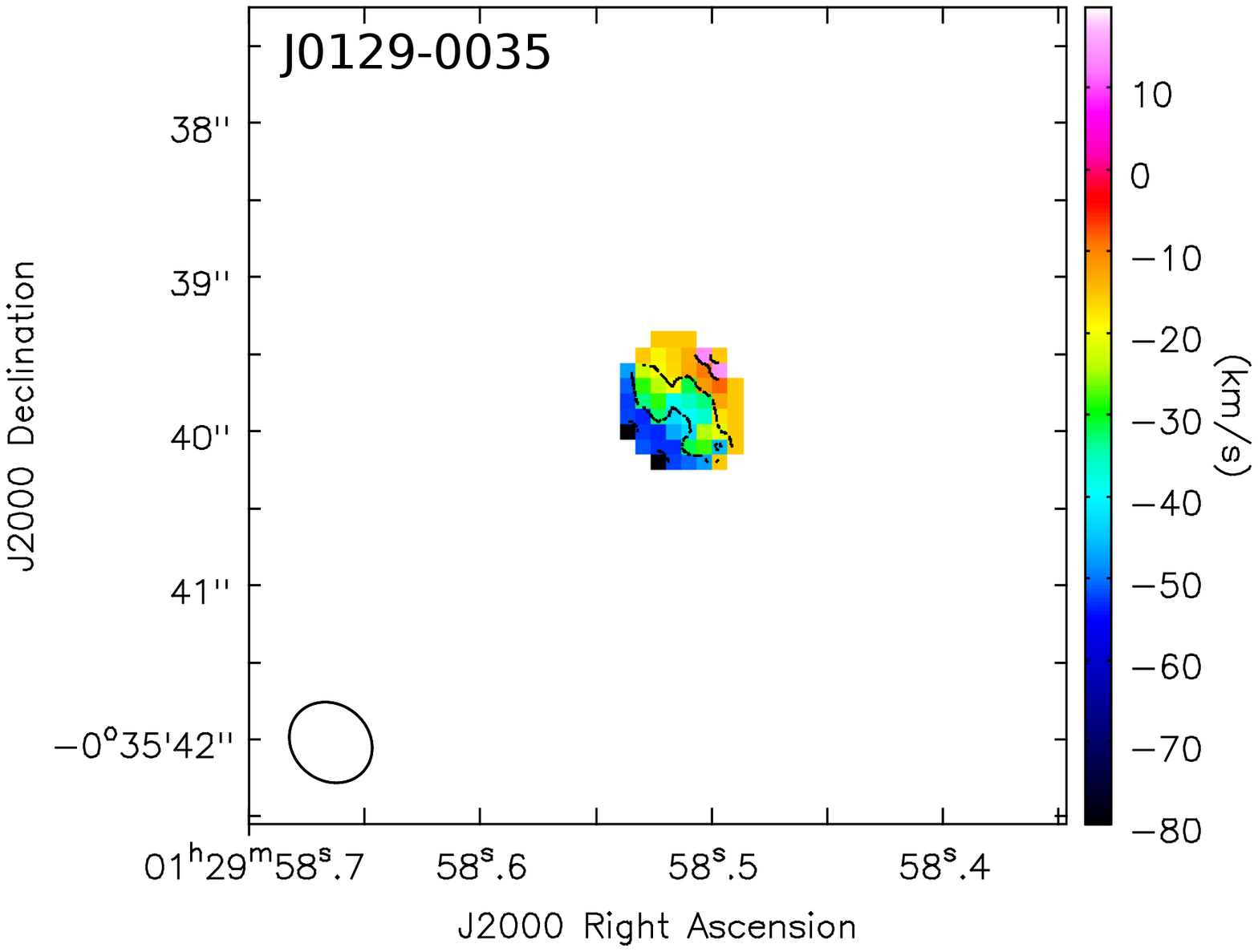}\\
\includegraphics[height=1.75in]{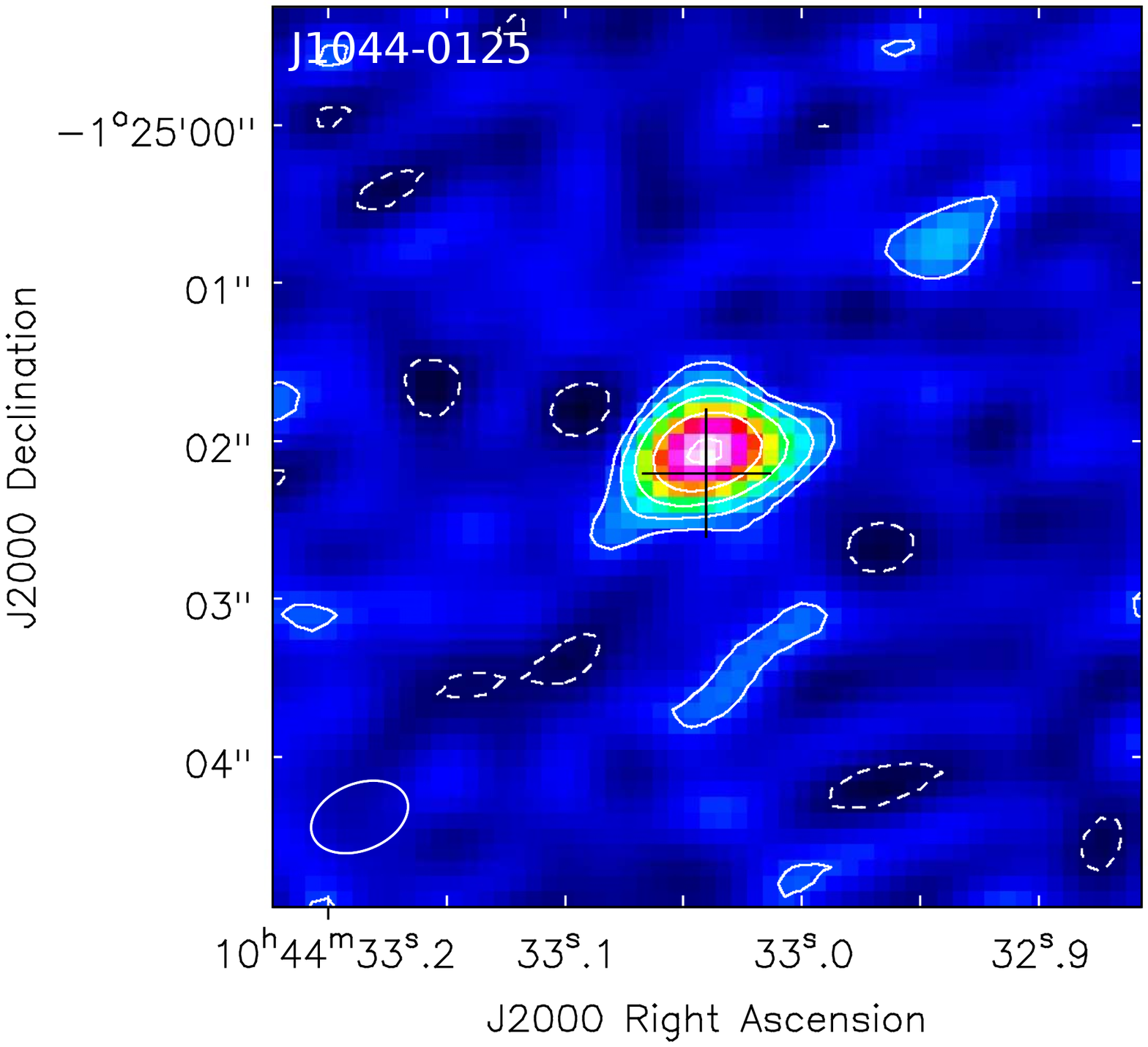}
\vskip -1.75in
\hspace*{2.0in}
\includegraphics[height=1.75in]{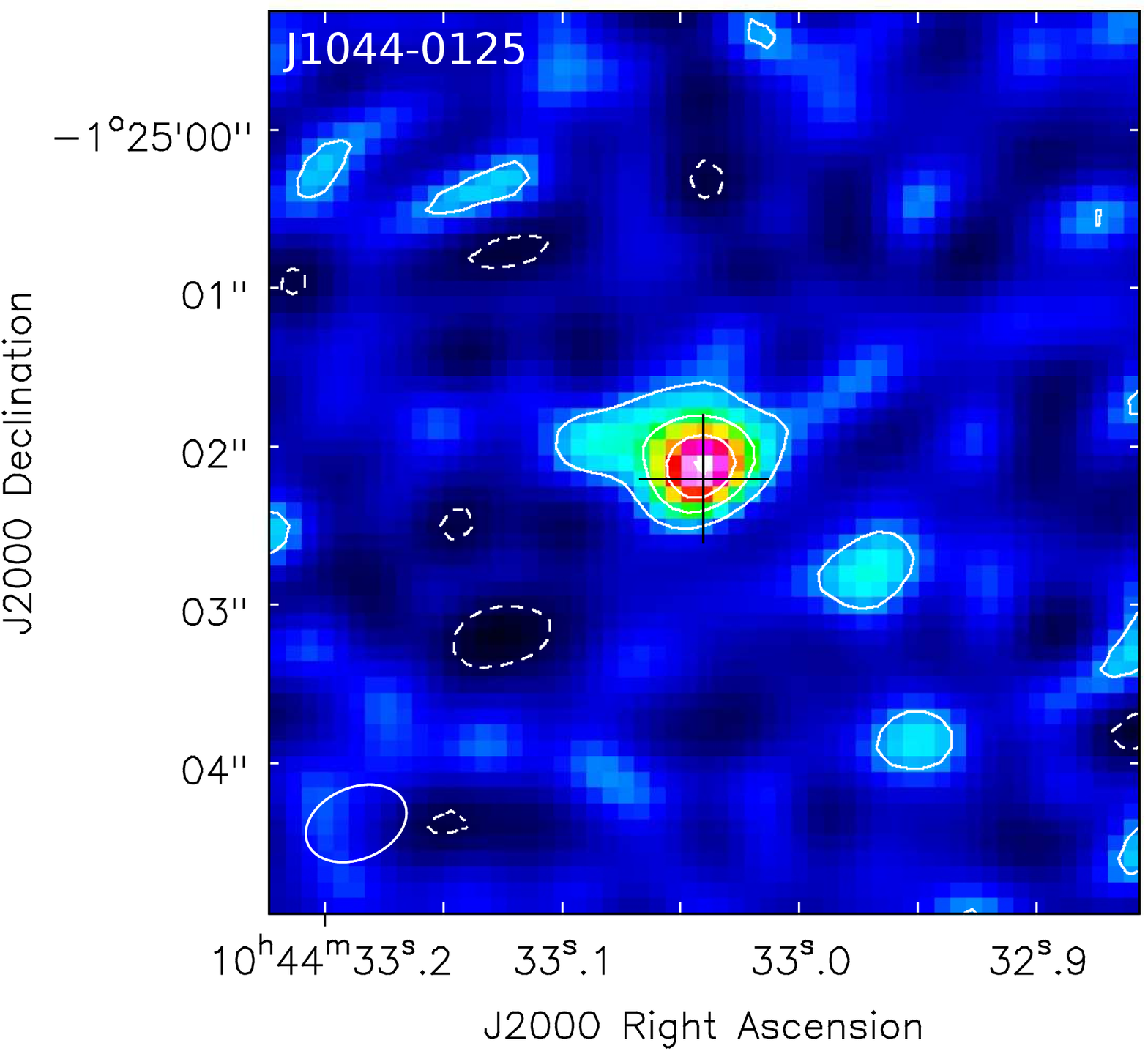}
\vskip -1.75in
\hspace*{4.0in}
\includegraphics[height=1.75in]{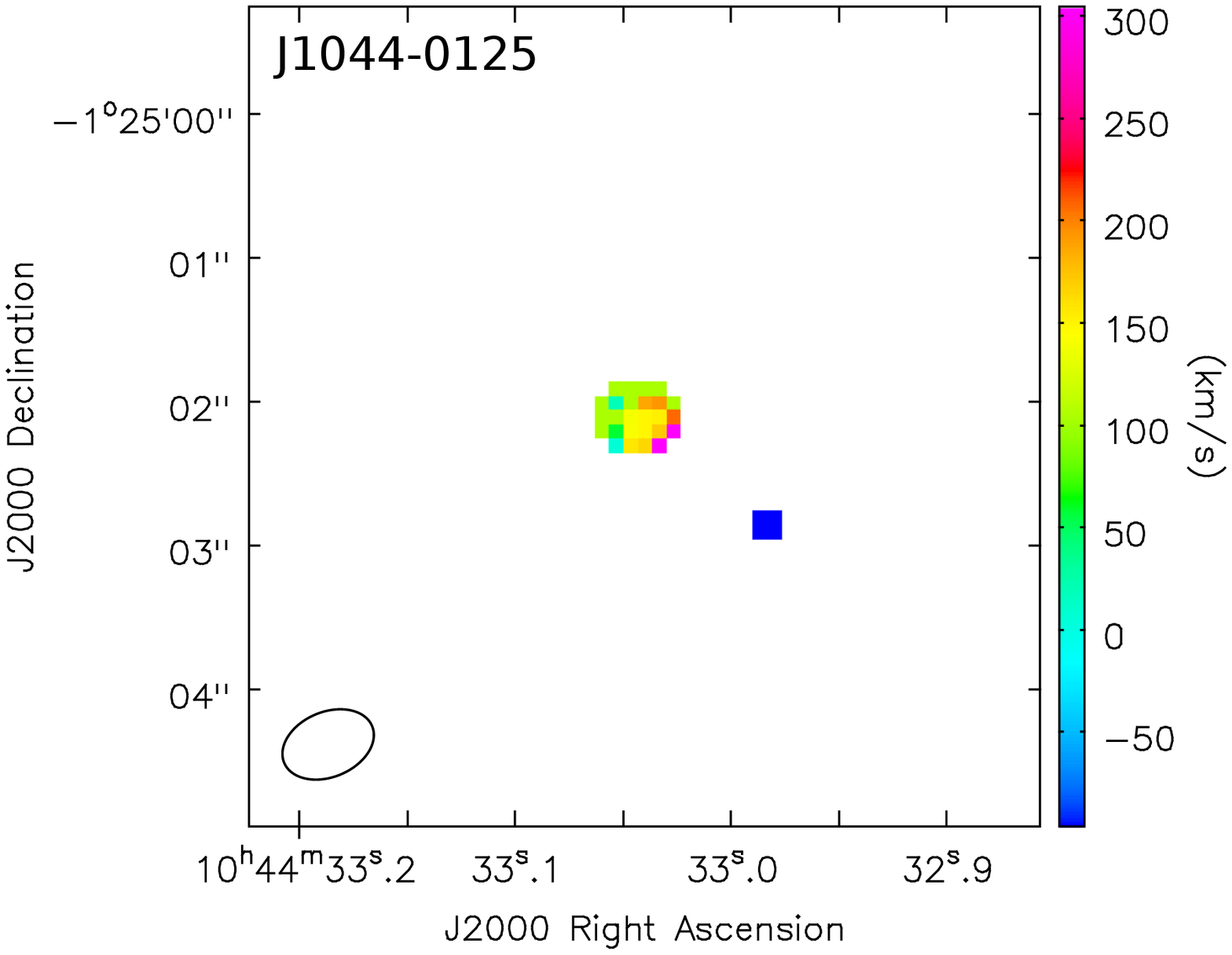}
\caption{Continued}
\end{figure}
\begin{figure}[h]
\plotone{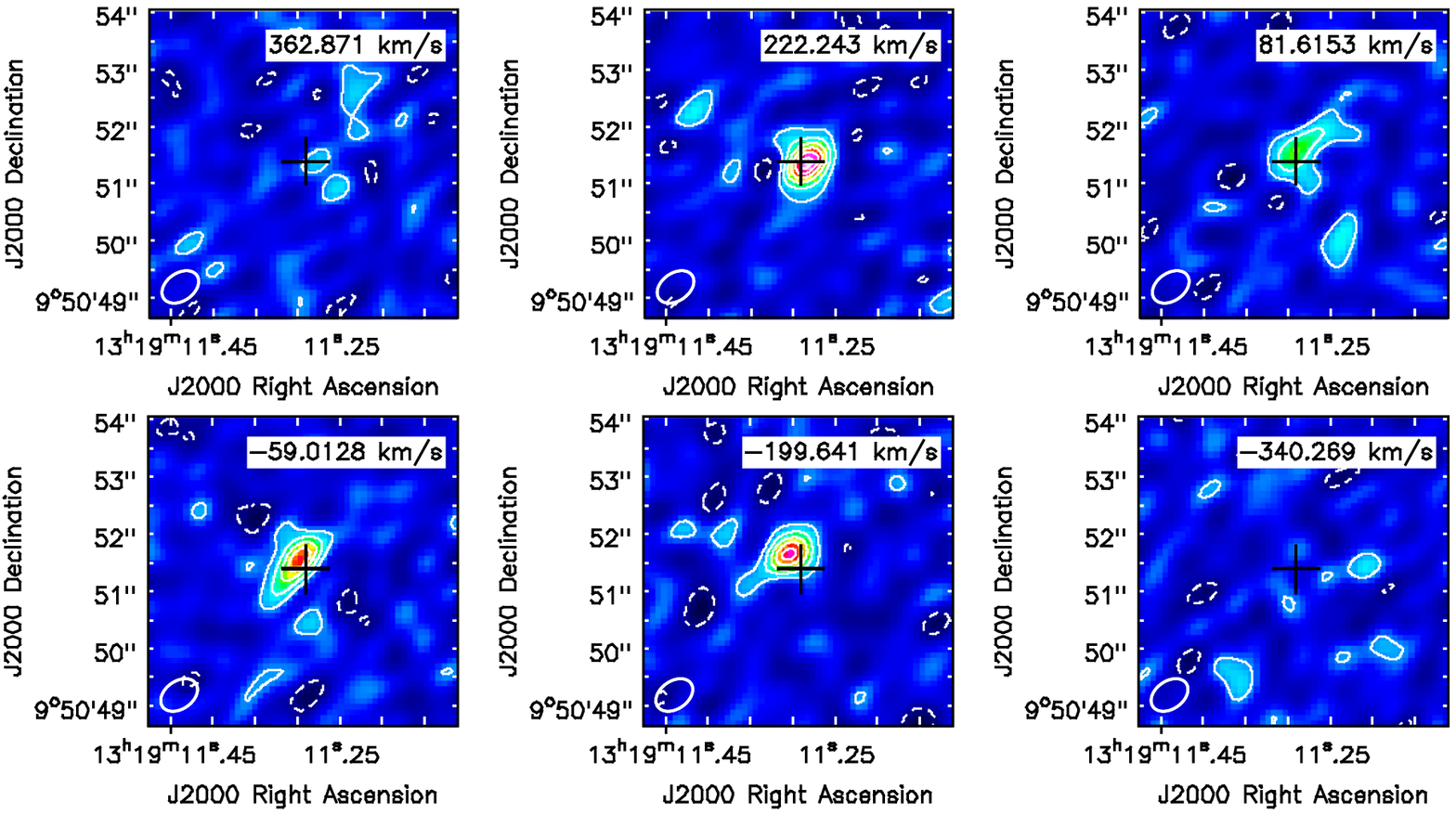}
\caption{\small [C II] line channel maps of J1319+0950 in the velocity range of $\rm -340.3$
to $\rm +362.9\,km\,s^{-1}$ at $\rm 0.7''\times0.5''$ resolution.
The channel width is $\rm 140\,km\,s^{-1}$ for each map, and the line is clearly detected in
the middle four channels. The typical $\rm 1\sigma$ rms noise of the channel map 
is $\rm 0.54\,mJy\,Beam^{-1}$, and the contours are 
[-2, 2, 4, 6, 8, 10, 12]$\rm \times 1\sigma$. 
The peak of the [C II] line emission is shifted by about 0.4$''$ from 
the $\rm 222\,km\,s^{-1}$ channel to the $\rm -200\,km\,s^{-1}$ channel. 
The cross denotes the location of the optical quasar.}
\end{figure}
\begin{figure}[h]
\epsscale{0.7}
\plotone{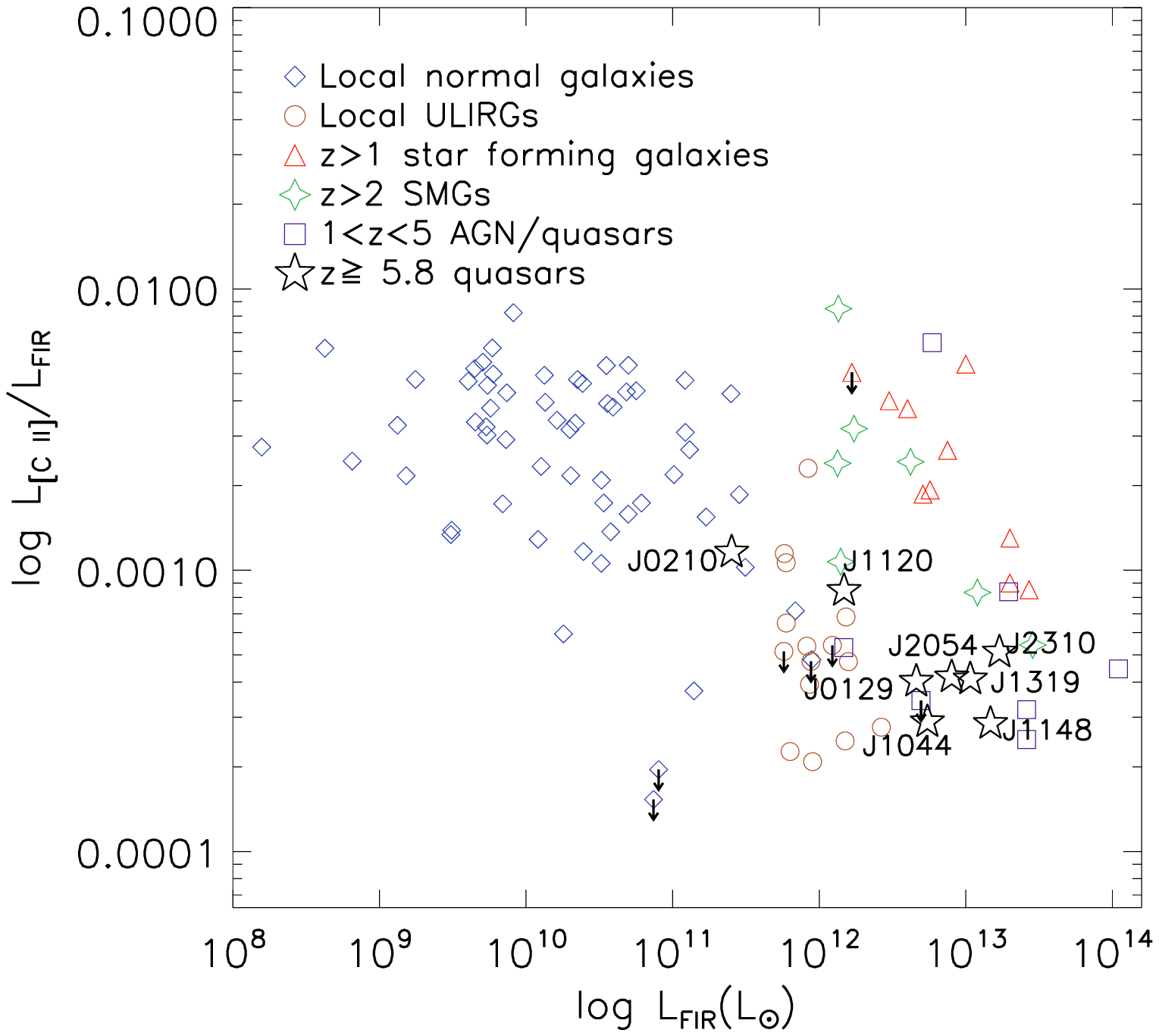}
\caption{\small $\rm L_{[CII]}/L_{FIR}$ vs. $\rm L_{FIR}$. 
The black stars indicate the quasars at z$\geq$5.8, including 
the five new [C II]-detected z$\sim$6 quasars in this work, 
J1148+5251 at z=6.42 \citep{maiolino05,leipski13}, CFHQS J0210$-$0456 at z=6.43 \citep{willott13}, and the z=7.08 
quasar ULAS J112001.48+064124.3 \citep{venemans12}.
We also plot the luminosity ratios from samples of 
local normal star forming galaxies \citep{malhotra01}, 
ULIRGs \citep{luhman03}, $z>1$ star forming galaxies and mixed systems \citep{stacey10,marsden05}, 
$z>2$ submillimeter bright galaxies (SMG, \citealp{maiolino09,ivison10,de11,swinbank12,wagg12,valtchanov11,riechers13}), 
and $1<z<5$ [C II] detected quasars \citep{pety04,maiolino05,maiolino09,
gallerani12,wagg12,carilli13} for comparison.  
}
\end{figure}
\begin{figure}[h]
\includegraphics[height=1.7in]{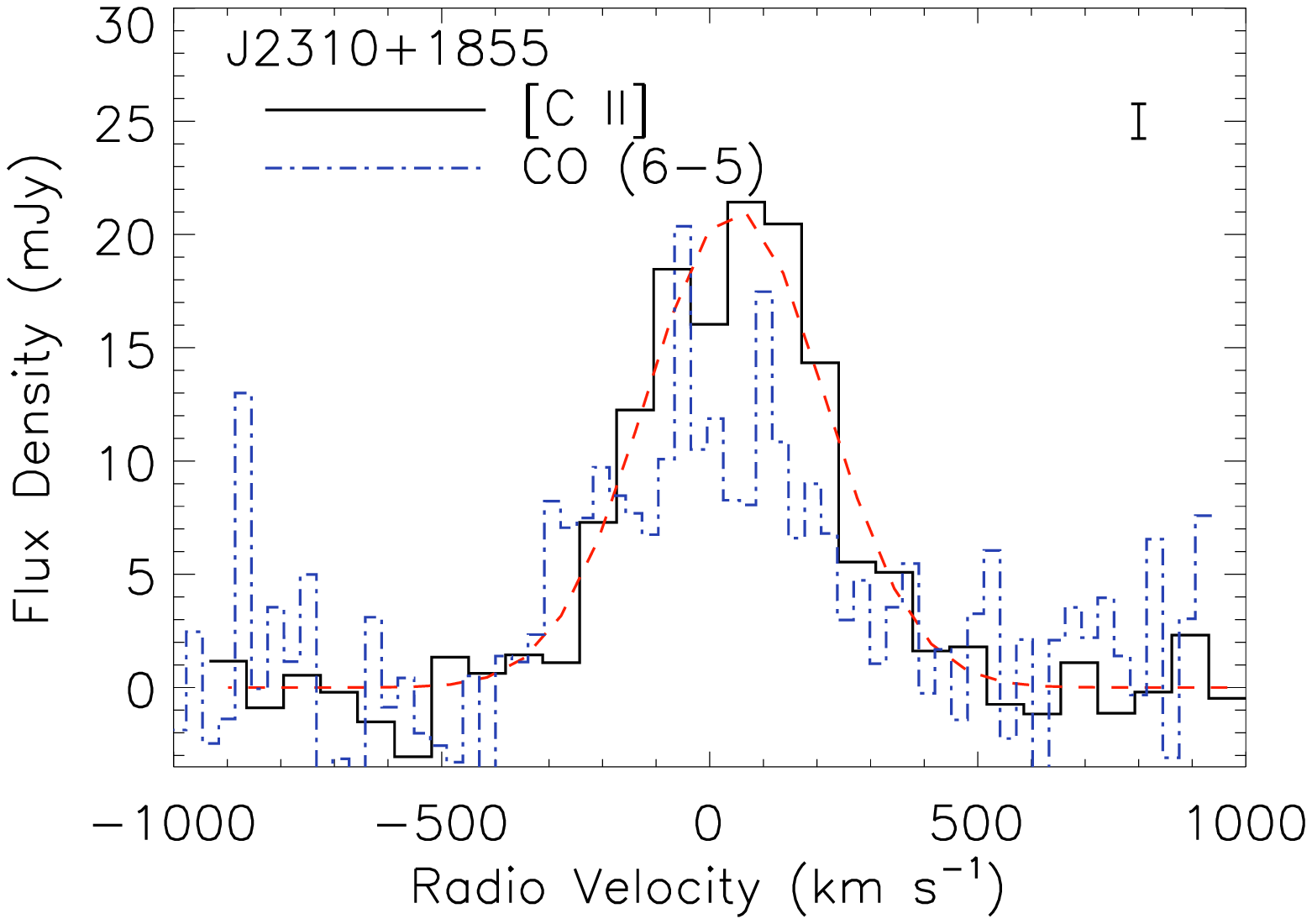}
\vskip -1.7in
\hspace*{2.1in}
\includegraphics[height=1.7in]{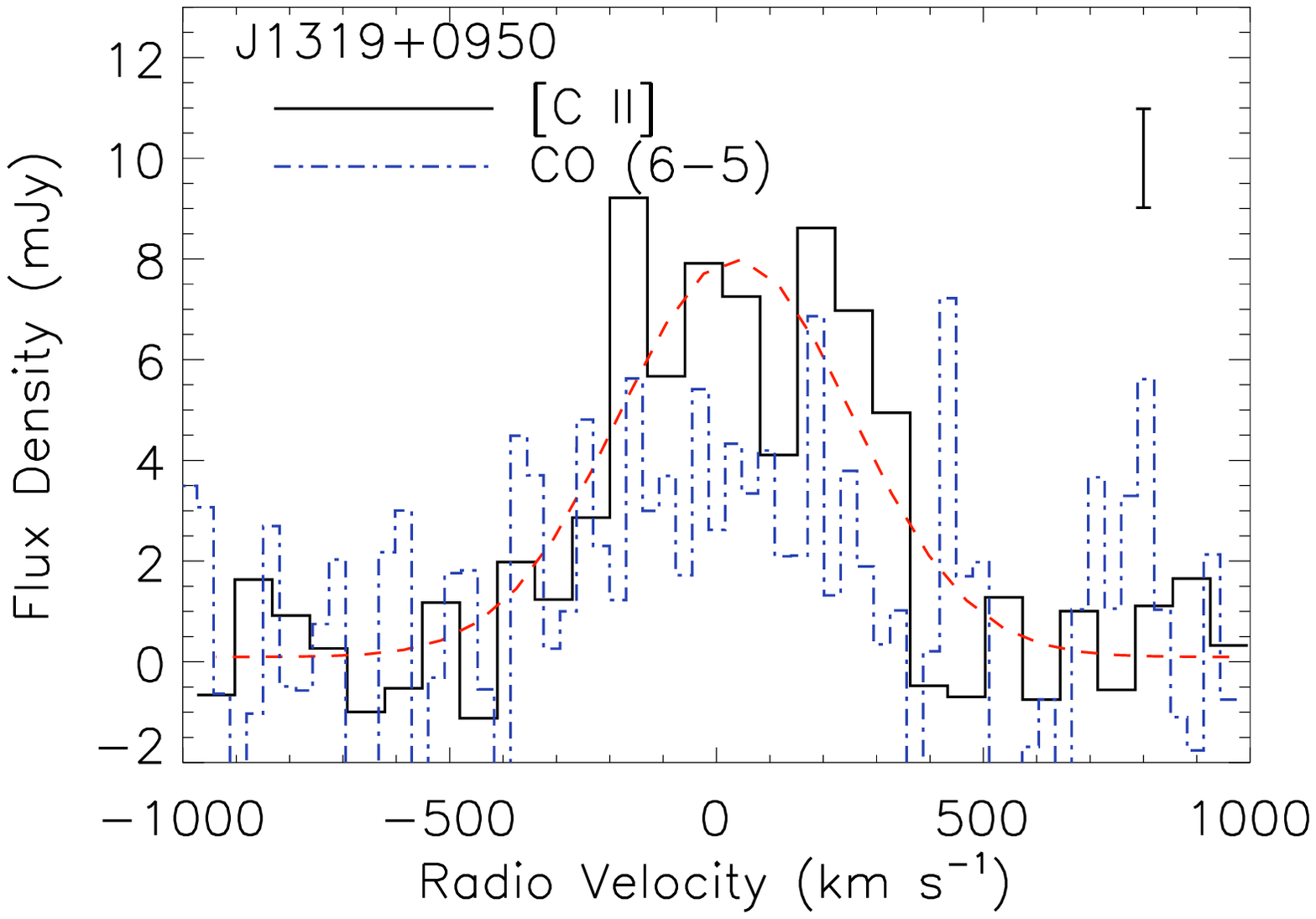}
\vskip -1.7in
\hspace*{4.2in}
\includegraphics[height=1.7in]{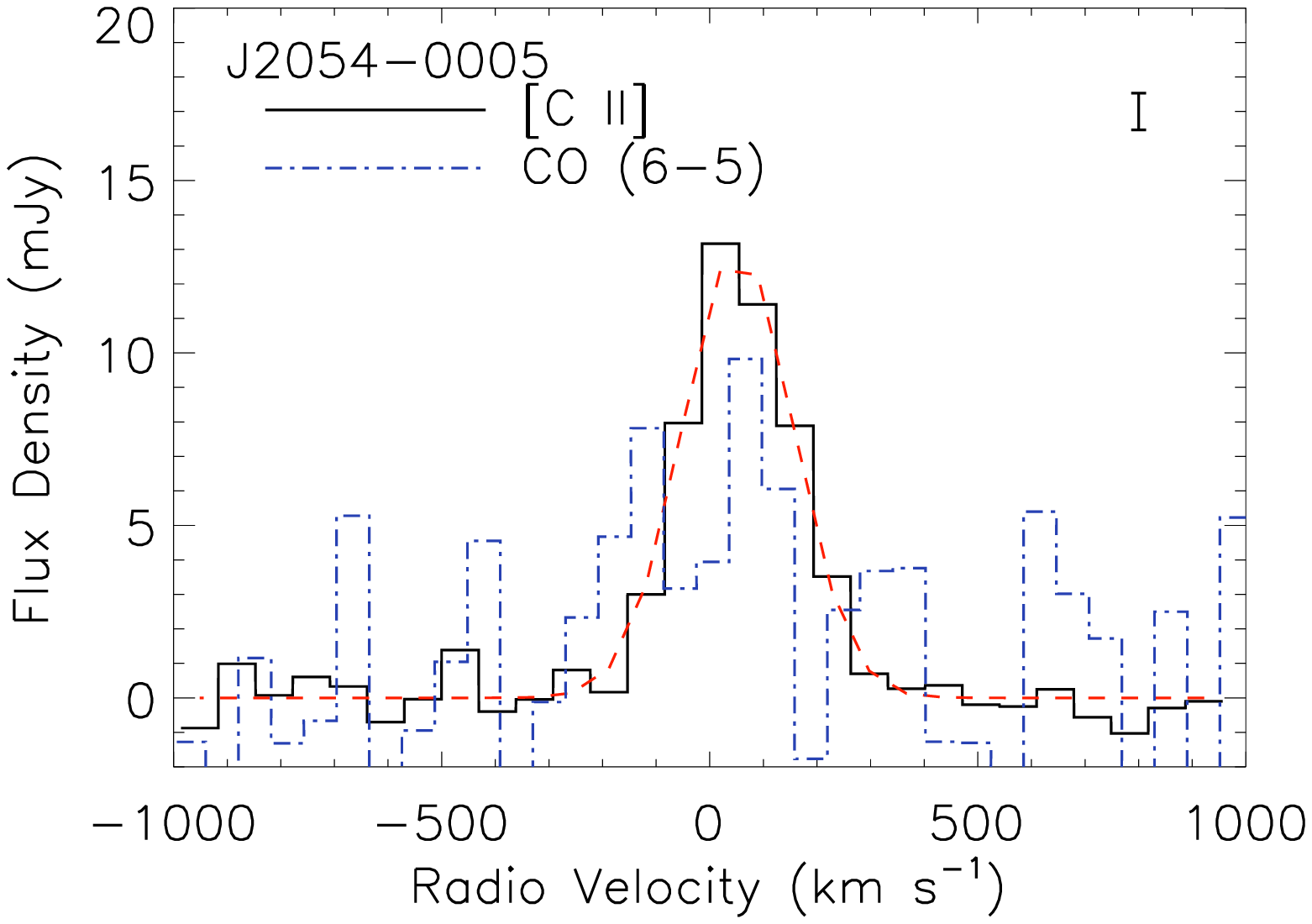}\\
\hspace*{1.0in}
\includegraphics[height=1.7in]{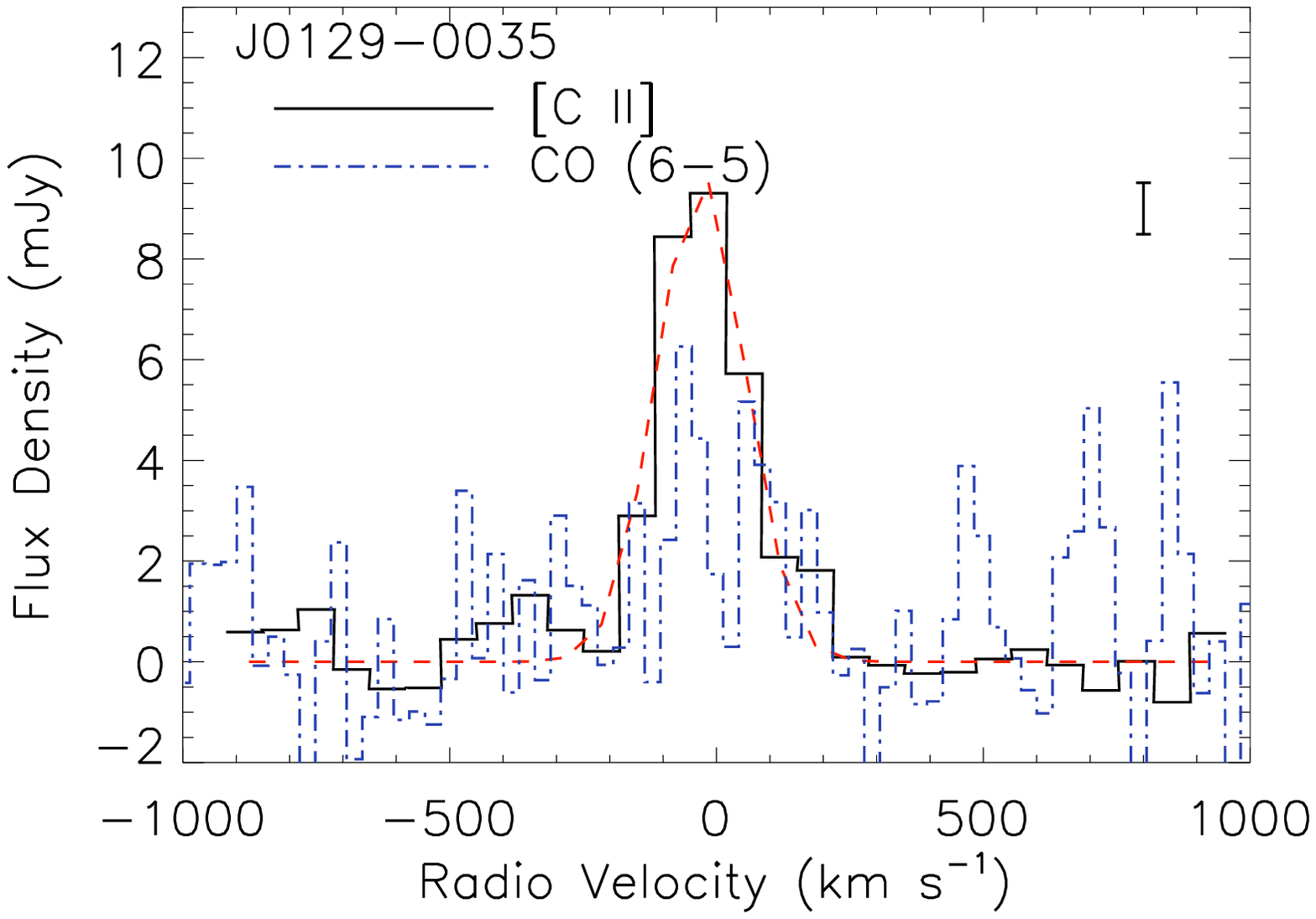}
\vskip -1.7in
\hspace*{3.2in}
\includegraphics[height=1.7in]{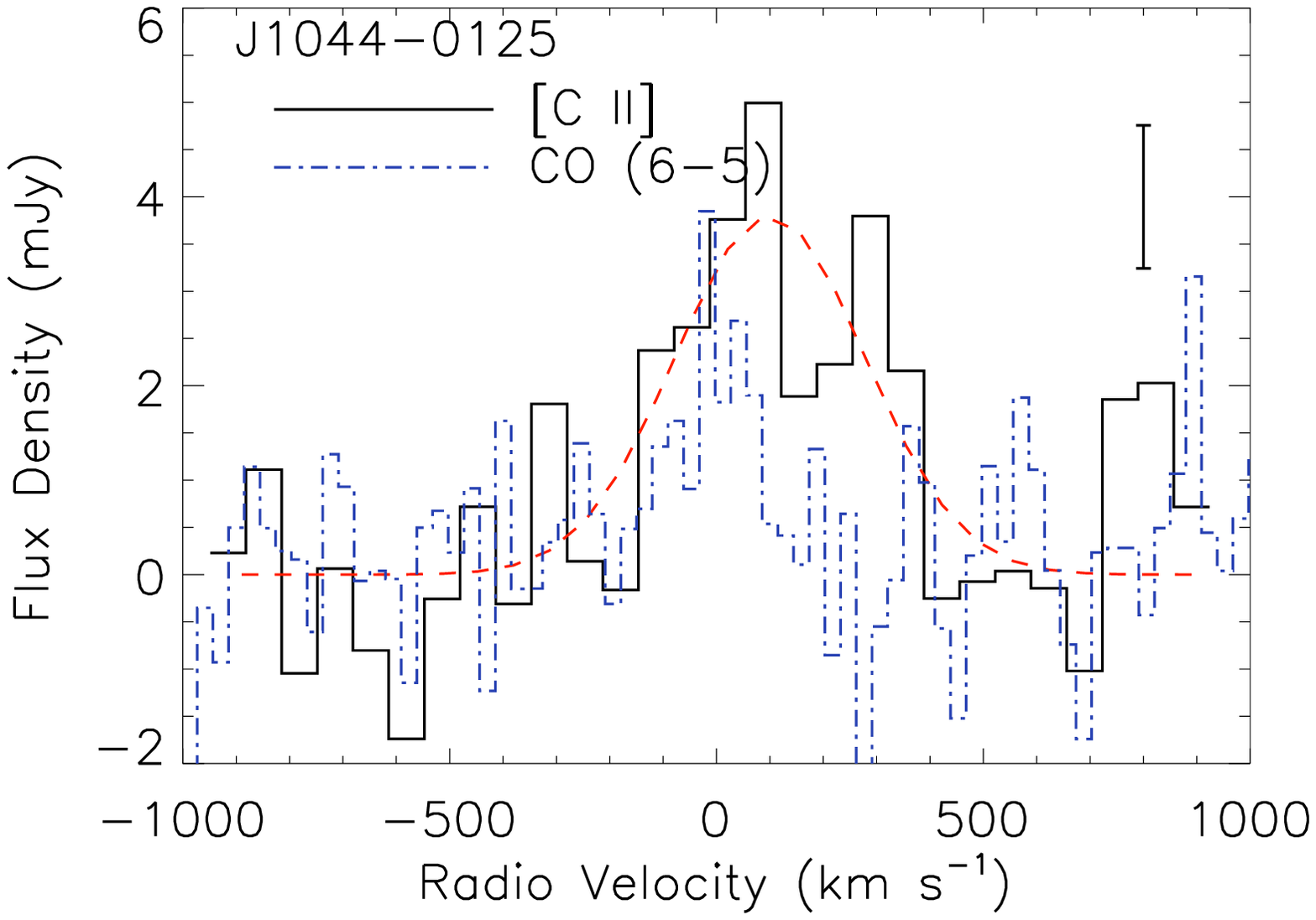}
\caption{\small The [C II] line spectra (black solid line) of the five ALMA 
detected quasars integrated over the line-emitting region in each channel, together with the previous CO (6-5)
detections from PdBI (blue dotted line, scaled to the [C II] line) and a
Gaussian fit to the [C II] line (red dashed line). 
The typical $\rm \pm$1$\sigma$ error per
channel for the [C II] line spectrum is shown in the top right of each panel. 
The zero velocity corresponds to the CO redshift listed in Table 1.}
\end{figure}




\end{document}